\documentclass{article}

\usepackage{graphicx}
\usepackage{amsmath}
\usepackage{amssymb}
\usepackage{amsthm}
\usepackage{mathtools, xcolor}

\usepackage[margin=2.5cm]{geometry}
\usepackage[T1]{fontenc}
\usepackage{lmodern}
\usepackage{microtype}
\usepackage{tikz}
\usetikzlibrary{arrows.meta,positioning,fit,backgrounds}

\title{Can a self-sustaining chemical network emerge due to just a few catalysts?}

\author{
Jinsu Kim\thanks{Department of Mathematics, Pohang University of Science and Technology (POSTECH), Pohang, Republic of Korea}
\and
Ercan Sönmez\thanks{Department of Mathematics, Pohang University of Science and Technology (POSTECH), Pohang, Republic of Korea}
}

\date{}

\usepackage{hyperref}

\newtheorem{theorem}{Theorem}
\newtheorem{proposition}[theorem]{Proposition}
\newtheorem{lemma}[theorem]{Lemma}
\newtheorem{corollary}[theorem]{Corollary}

\theoremstyle{definition}
\newtheorem{definition}[theorem]{Definition}
\newtheorem{assumption}[theorem]{Assumption}

\theoremstyle{remark}
\newtheorem{remark}[theorem]{Remark}

\newcommand{\dln}{\displaystyle \lim_{n\to\infty}}

\begin{document}

\maketitle

\begin{abstract}
Understanding how self-sustaining chemical reaction networks can emerge from simple molecules is an important question in chemical approaches to the origin of life. Reflexively autocatalytic and food-generated (RAF) sets provide a mathematical framework for investigating this question. This line of research goes back to the early work of Kauffman on autocatalytic reaction networks and was subsequently developed through the probabilistic framework of Mossel and Steel. Further contributions by Hordijk, Steel, and collaborators extended the theory to more heterogeneous forms of catalysis, including reaction-dependent probabilities or power-law distributions of catalytic activity. We introduce a stochastic model that combines heterogeneous molecular catalytic activity with reaction-dependent molecule–reaction compatibility. By assigning random catalytic weights to molecules and incorporating compatibility through an explicit formulation of catalysis probabilities, our model provides a concrete and flexible setting that captures both sources of heterogeneity while remaining mathematically tractable.

We establish novel and general criteria for RAF formation and nonformation. We further demonstrate that classical phase-transition criteria need not remain valid in the presence of reaction-dependent catalysis. For heavy-tailed catalytic weights, we identify and rigorously quantify an additional phase transition. In models with finite-mean catalytic weights, RAF formation and nonformation are governed by a catalytic intensity parameter. For infinite-mean weights, however, we show that RAFs may emerge with probability tending to one even when the catalytic intensity parameter tends to zero.
\end{abstract}


\medskip


\tableofcontents

\section{Introduction}
A self-sustaining chemical reaction system is one in which molecules are produced from available resources through reactions that collectively support one another. Such systems have attracted considerable attention because they may help explain how organized chemical systems emerged during the origin of life. An interesting mathematical question is how likely a self-sustaining system is to arise when the catalysis network is random. That is, for a fixed set of possible reactions, if the catalytic relationships between molecules and reactions are randomly determined, how likely is it that a self-sustaining system exists? This question has been studied in a series of pioneering works by Mossel, Steel, Hordijk, and their collaborators, who developed mathematical frameworks for studying the emergence of reflexively autocatalytic and food-generated (RAF) sets. In particular, Mossel and Steel~\cite{MosselSteel2005} established fundamental conditions under which RAFs emerge with high probability in random polymer models. Our study extends this line of research by introducing random molecular weights that represent differences in catalytic activity among molecules, while allowing the probability of catalysis to depend on both the molecule and the reaction. In this setting, we provide new insights into the possibility that self-sustaining systems can emerge even when catalytic activity is concentrated in a small number of molecules.

The emergence of self-sustaining chemical reaction networks has long been considered a potentially important step in the origin of life. Kauffman~\cite{Kauffman1986} proposed that sufficiently complex molecular systems could give rise to collectively autocatalytic networks, an idea subsequently explored in polymer-based models of chemical reactions. Experimental studies have demonstrated self-sustained replication and cooperative catalytic networks in RNA systems. More recent research has investigated autocatalysis in connection with molecular replication, catalytic regulation, spatial organization, and the organization of prebiotic chemical systems \cite{lu2024small,kriukov2024exploring,PlumEtAl2025,
SakrefRivoire2024,ZorcRoy2024,Despons2025}. These developments motivate the mathematical investigation
of the conditions under which autocatalytic networks can emerge. 

Among the many previous studies, several papers are particularly relevant to our work. Mossel and Steel~\cite{MosselSteel2005} provided a rigorous framework for random catalytic networks with independent
catalysis events and reaction-independent catalysis probabilities for each molecule. They established fundamental bounds on the amount of catalysis required for RAF formation as the maximum polymer length increases. Hordijk, Wills and Steel~\cite{HordijkWillsSteel2014} extended the investigation to reaction-dependent catalysis mechanisms in the binary polymer model. Hordijk and Steel~\cite{HordijkSteel2016} studied heterogeneous catalytic activity with each molecule assigned to  a random number of reactions that it catalyzes, while the reactions are selected uniformly. They showed that the distribution of catalytic activity can significantly affect the probability and size of RAFs, while universal linear order bounds on the average catalytic activity required for RAF formation continue to hold. Phase transitions in RAF probability with respect to the catalytic probability was also studied in Varanasi and Korenaga \cite{varanasi2026emergence}. See Section \ref{sec:comparison} for more details about comparison to previous studies.

The main structure of the model used in this paper follows the frameworks developed by Mossel and Steel \cite{MosselSteel2005} and Hordijk and Steel \cite{HordijkSteel2016}. In the reaction-independent models considered in \cite{MosselSteel2005,HordijkSteel2016}, the probability that a molecule catalyzes a reaction depends on the molecule but not on the particular reaction. We extend this framework by considering intensity parameters, which determine the probability of catalysis and depend on both the catalyst and the reaction. We also introduce random molecular weights to represent heterogeneity in catalytic activity among molecules. Our extension is analogous to the extension from the Erd\"os--R\'enyi random graph model to inhomogeneous random graph models, where the probability of an edge depends on the properties of the vertices. In our setting, the molecules and reactions form two classes of vertices, and their catalytic relationships are determined by reaction-dependent intensities and molecular weights.

Our main results provide new phase-transition criteria for the probability of existence of a RAF in terms of the catalytic intensities and molecular weights. In particular, we identify regimes in which $\mathbb{P}(\mathsf{RAF})\to 1$ and regimes in which $\mathbb{P}(\mathsf{RAF})\to 0$ as the maximum polymer length $n$ tends to infinity (see Theorem \ref{thm:weighted-mass-criterion}). These results reveal phenomena that are qualitatively different from those characterized by earlier models. In particular, Mossel and Steel~\cite{MosselSteel2005} established fundamental criteria for RAF formation in terms of the expected number of reactions catalyzed by individual molecules. Our results extend their analysis to a more general setting by introducing random molecular weights and allowing catalysis probabilities to depend on both molecules and reactions. 

One particularly interesting consequence is that a RAF can emerge with probability tending to one even in regimes where catalytic activity is concentrated in a small number of molecules (see Figure \ref{fig:raf samples}). This can happen even with a single molecule (see Proposition \ref{prop:single-catalyst-raf}). This consequence is consistent with the numerical simulations shown in Hordijk and Steel \cite{HordijkSteel2016}. In particular, heavy-tailed molecular weights allow a few molecules to have substantially greater catalytic influence than the others, while the overall intensity of catalysis may decrease with the system size. Nevertheless, sufficiently strong and appropriately distributed catalytic interactions can ensure that
$\mathbb{P}(\mathsf{RAF})\to 1$ as $n\to\infty$. This suggests that the emergence of collectively autocatalytic systems does not necessarily require catalytic activity to be distributed broadly among molecules. Instead, a small number of highly active molecules may play a decisive role in supporting the emergence of a RAF. Although our model does not directly describe the chemical dynamics of early life, this phenomenon provides a possible mathematical perspective on how organized autocatalytic systems might emerge from heterogeneous molecular populations.

One of the main mathematical challenges is the dependence of catalysis probabilities on both catalysts and reactions, together with an additional source of randomness in the model arising from molecular weights. Our proofs of the RAF formation criteria follow the general constructive approach of Mossel and Steel~\cite{MosselSteel2005}, which was subsequently extended to reaction-dependent catalysis mechanisms by Hordijk, Wills and Steel~\cite{HordijkWillsSteel2014}. The key observation in our weighted model is that, conditional on the molecular weights, catalysis events are independent, and the probability that a given reaction $r$ receives no catalyst is exactly $\exp(-\lambda_n M_n(r))$, where $M_n(r)$ denotes its weighted catalytic mass (see \eqref{eq:molecular weights}). This allows us to formulate general criteria for RAF formation in terms of the asymptotic behavior of these masses. The resulting criteria are independent of the specific weight distribution, so that the subsequent analysis of heavy-tailed weights reduces to establishing the corresponding asymptotic properties of the catalytic masses.

The main novelty of our results lies in their formulation in terms of weighted catalytic masses, which provides general criteria for RAF formation that simultaneously accommodate reaction-specific catalysis and heterogeneous molecular catalytic activity. This framework allows us to identify and mathematically quantify phase transitions, in particular for heavy-tailed catalytic weights, and to distinguish different regimes of RAF formation. Beyond these phase-transition results, our analysis also reveals limitations of earlier criteria. In particular, we show that reaction-dependent catalysis alone already provides counterexamples to extending the Mossel--Steel criteria beyond their original assumptions. Thus, reaction specificity and molecular heterogeneity are not merely additional features of random catalytic models; they can fundamentally affect the probability of RAF formation.

\section{Model and main results}

We begin this section by introducing the fundamental definitions and notation used throughout the paper.

\subsection{Catalytic reaction systems and autocatalytic sets}
The catalytic reaction systems in this paper have the same format and structure as shown in \cite{MosselSteel2005}. 

Fix a finite alphabet \(\mathcal A=\{a_1,\ldots,a_\kappa\}\) with \(\kappa\ge 2\). For \(n\in\mathbb N\), let \(X_n:=\bigcup_{j=1}^{n}\mathcal A^j\)
be the set of all non-empty strings over \(\mathcal A\) of length at most \(n\). Its cardinality is \(|X_n|=\sum_{j=1}^{n}\kappa^j =\frac{\kappa^{n+1}-\kappa}{\kappa-1}.\)
For \(x\in X_n\), write \(|x|\) for its length (or size). For example, if $x=a_1a_2$, then $|x|=2$. 
\paragraph{The deterministic reaction system.}
Whenever \(u,v\in X_n\) satisfy \(|u|+|v|\le n\), we allow the ligation reaction \(u+v\longrightarrow uv,\)
where \(uv\) denotes concatenation. The set of forward ligation reactions is \(R_n^+ := \left\{ (u,v;uv): u,v\in X_n,\quad |u|+|v|\le n \right\}.\)
Associated with each ligation reaction is the reverse cleavage reaction \(uv\longrightarrow u+v.\)
Accordingly, define
\[
R_n^- := \left\{ (uv;u,v): (u,v;uv)\in R_n^+ \right\}, \qquad R_n:=R_n^+\cup R_n^-.
\]
Forward and reverse reactions are regarded as distinct directed reactions. When only ligation reactions are required, we restrict attention to \(R_n^+\).
For each \(r\in R_n\), let \(\rho(r)\subseteq X_n\) and \(\pi(r)\subseteq X_n\) denote its sets of reactants and products, respectively. Thus
\[
\begin{aligned} \rho((u,v;uv))&=\{u,v\}, & \pi((u,v;uv))&=\{uv\}, \\ \rho((uv;u,v))&=\{uv\}, & \pi((uv;u,v))&=\{u,v\}.
\end{aligned}
\]
These sets record molecule types rather than stoichiometric multiplicities; the generation conditions below concern the availability of these types.
\paragraph{Food set and \(F\)-generation.}
Fix an integer \(t\ge 1\), independently of \(n\), and henceforth assume \(n>t\). Define the food set by
\[
F_n:=\{x\in X_n:|x|\le t\}=X_t. 
\]
In particular, \(F_n\) is independent of \(n\) for \(n>t\).
For \(R'\subseteq R_n\), define the closure of \(F_n\) under \(R'\), denoted \(\operatorname{cl}_{R'}(F_n)\), to be the smallest subset \(Y\subseteq X_n\) satisfying \(F_n\subseteq Y\)
and
\[
\rho(r)\subseteq Y \quad\Longrightarrow\quad \pi(r)\subseteq Y \qquad \text{for every }r\in R'.
\]
Equivalently, \(\operatorname{cl}_{R'}(F_n)\) is the collection of molecules that can be generated from the food set by successively applying reactions from \(R'\). 
A reaction set \(R'\subseteq R_n\) is called \emph{\(F\)-generated} if
\[
\rho(r)\subseteq\operatorname{cl}_{R'}(F_n) \qquad \text{for every }r\in R'.
\]
\paragraph{Catalysis configurations.}
A catalysis configuration is a subset \(C_n\subseteq X_n\times R_n.\)
We interpret \((x,r)\in C_n\) as the statement that molecule \(x\) catalyses reaction \(r\). Equivalently, introduce the indicators \(I_{x,r}:=\mathbf 1_{\{(x,r)\in C_n\}}, \qquad x\in X_n,\quad r\in R_n.\)
The definitions below apply to any catalysis configuration; its probability law will be specified subsequently.
\paragraph{Reflexively autocatalytic and food-generated (RAF) sets.}
Given a catalysis configuration \(C_n\), a non-empty reaction set \(R'\subseteq R_n\) is called \emph{reflexively autocatalytic and food-generated}, abbreviated \emph{RAF}, if the following two conditions hold.
\begin{enumerate}
\item \emph{\(F\)-generation.}
Every reactant required by \(R'\) can be generated from the food set:
\[
\rho(r)\subseteq\operatorname{cl}_{R'}(F_n) \qquad \text{for all }r\in R'.
\]
\item \emph{Reflexive autocatalysis.}
Every reaction in \(R'\) is catalysed by at least one molecule available from the food set or generated by \(R'\):
\[
\forall r\in R', \qquad \exists x\in\operatorname{cl}_{R'}(F_n) \quad\text{such that}\quad I_{x,r}=1.
\]
\end{enumerate}
When the catalysis configuration is random, we write \(\mathsf{RAF}_n\) for the event that the resulting catalytic reaction system contains at least one RAF. For a system restricted to forward reactions, the same definitions apply with \(R_n\) replaced by \(R_n^+\). See Figure \ref{fig:raf-vs-nonraf} for examples of a RAF set and a non-RAF set.

\begin{figure}[htbp]
\centering

\tikzset{
  molecule/.style={
    circle,draw=black!75,thick,
    minimum size=9mm,inner sep=1pt
  },
  food/.style={
    molecule,fill=yellow!12
  },
  formed/.style={
    molecule,fill=green!12
  },
  absent/.style={
    molecule,fill=gray!12,
    draw=gray!75!black
  },
  reaction/.style={
    rectangle,rounded corners=1.5pt,
    draw=black!75,thick,
    minimum width=10mm,
    minimum height=7mm,
    fill=white
  },
  reactionarrow/.style={
    -{Stealth[length=2mm]},thick
  },
  catalytic/.style={
    -{Stealth[length=2mm]},
    dashed,thick,orange!85!black
  }
}

\begin{minipage}[t]{0.48\textwidth}
\centering
\textbf{(i) RAF}\par\medskip

\resizebox{\linewidth}{!}{%
\begin{tikzpicture}[font=\small]

  \node[food] (a) at (0,1.4) {$a$};
  \node[food] (b) at (0,0) {$b$};

  \node[reaction] (r1) at (2,1.15) {$r_1$};
  \node[formed] (ab) at (4,1.15) {$ab$};

  \node[reaction] (r2) at (6,1.15) {$r_2$};
  \node[formed] (abb) at (8,1.15) {$abb$};

  \draw[reactionarrow] (a) -- (r1);
  \draw[reactionarrow] (b) -- (r1);
  \draw[reactionarrow] (r1) -- (ab);

  \draw[reactionarrow] (ab) -- (r2);
  \draw[reactionarrow]
    (b.south) |- (5,-0.65) -| (r2.south);
  \draw[reactionarrow] (r2) -- (abb);

  \draw[catalytic]
    (abb.north) |- (5,2.55) -| (r1.north);

  \draw[catalytic]
    (ab.south)
    to[out=-55,in=-125] (r2.south);

  \node[
    anchor=west,
    font=\scriptsize,
    orange!85!black
  ] at (5.05,2.75) {catalysis};

\end{tikzpicture}%
}

\vspace{2mm}

{\footnotesize
$F=\{a,b\}$, $R'=\{r_1,r_2\}$,

$r_1:a+b\to ab$,

$r_2:ab+b\to abb$.

\medskip
$\operatorname{cl}_{R'}(F)
=\{a,b,ab,abb\}$.
\par}

\end{minipage}
\hfill
\begin{minipage}[t]{0.48\textwidth}
\centering
\textbf{(ii) Not a RAF}\par\medskip

\resizebox{\linewidth}{!}{%
\begin{tikzpicture}[font=\small]

  \node[food] (a) at (0,1.4) {$a$};
  \node[food] (b) at (0,0) {$b$};

  \node[reaction] (r1) at (2,1.15) {$r_1$};
  \node[formed] (ab) at (4,1.15) {$ab$};

  \node[reaction] (r2) at (6,1.15) {$r_2$};
  \node[absent] (abc) at (8,1.15) {$abc$};
  \node[absent] (c) at (5,-0.45) {$c$};

  \draw[reactionarrow] (a) -- (r1);
  \draw[reactionarrow] (b) -- (r1);
  \draw[reactionarrow] (r1) -- (ab);

  \draw[reactionarrow] (ab) -- (r2);
  \draw[reactionarrow] (c) -- (r2);
  \draw[reactionarrow] (r2) -- (abc);

  \draw[catalytic]
    (ab.north) |- (3,2.55) -| (r1.north);

  \draw[catalytic]
    (c.east)
    to[out=5,in=-90,looseness=1.1]
    (r2.south east);

  \node[
    anchor=west,
    font=\scriptsize,
    red!75!black
  ] at (6.6,-0.4)
    {$c\notin \operatorname{cl}_{R'}(F)$};

\end{tikzpicture}%
}

\vspace{2mm}

{\footnotesize
$F=\{a,b\}$, $R'=\{r_1,r_2\}$,

$r_1:a+b\to ab$,

$r_2:ab+c\to abc$.

\medskip
$\operatorname{cl}_{R'}(F)
=\{a,b,ab\}$.
\par}

\end{minipage}

\caption{
Examples of a RAF and a non-RAF reaction set
in a binary polymer system.
Solid black arrows indicate reactions,
while orange dashed arrows indicate catalysis.
(i) $R'=\{r_1,r_2\}$ is a RAF:
both reactions are $F$-generated and catalyzed
within $\operatorname{cl}_{R'}(F)$.
(ii) $R'=\{r_1,r_2\}$ is not a RAF:
the reactant $c$ of $r_2$ is not in
$\operatorname{cl}_{R'}(F)$,
so $R'$ is not $F$-generated.
Moreover, $r_2$ is catalyzed only by $c$,
which is also outside the closure.
Hence both RAF conditions fail.
}
\label{fig:raf-vs-nonraf}

\end{figure}

\subsection{A random catalysis model with heterogeneous molecular activity}

We now introduce a new model of random catalysis. To account for heterogeneity in catalytic activity, associate with each molecule \(x\in X_n\) a random weight \(W_x\). Assume that \(\{W_x:x\in X_n\}\) are independent and identically distributed copies of a random variable \(W\) satisfying \(\mathbb P(0<W<\infty)=1.\)
The weight \(W_x\) represents the intrinsic catalytic activity of molecule \(x\). No moment assumptions on \(W\) are imposed at this stage.
Let \(\lambda_n\in[0,\infty)\) be a deterministic intensity parameter, and let \(K_n:X_n\times R_n\longrightarrow[0,\infty)\)
be a deterministic kernel taking finite values. The quantity \(K_n(x,r)\) describes the compatibility of molecule \(x\) with reaction \(r\). Define the sigma-algebra generated by the molecular weights as \(\mathcal W_n:=\sigma(W_x:x\in X_n).\)
Conditional on \(\mathcal W_n\), the catalysis indicators \(\{I_{x,r}:(x,r)\in X_n\times R_n\}\) are mutually independent, with
\[
\mathbb P(I_{x,r}=1\mid\mathcal W_n) = 1-\exp\{-\lambda_nW_xK_n(x,r)\} \qquad\text{almost surely}.
\]

Note that our model relaxes both assumptions (R1) and (R2) of Mossel and Steel~\cite{MosselSteel2005}. Their assumption (R1) requires the catalysis indicators to be mutually independent. Assumption (R2) requires the probability that a molecule catalyses a forward reaction to depend only on the molecule $x$. Observe that both of these assumptions need not hold in our model. In particular independence of $I_{x,r}$ only hold conditioned on $\mathcal W_n$. 

Together with the law of the weights, the above conditional specification determines the joint law of the catalysis configuration \(C_n:=\{(x,r)\in X_n\times R_n:I_{x,r}=1\}.\)
The model separates three contributions to catalysis: the global intensity \(\lambda_n\), the intrinsic molecular activity \(W_x\), and the molecule--reaction compatibility \(K_n(x,r)\). In particular, \(K_n(x,r)=0\) excludes catalysis of \(r\) by \(x\), whereas larger values of \(W_xK_n(x,r)\) increase its conditional probability.
The independence assumption is conditional on the weights. Indicators associated with distinct molecules are also independent unconditionally, while those associated with the same molecule need not be, since they share the random weight \(W_x\).

\subsubsection{Comparison with previous models}\label{sec:comparison}

The polymer model underlying our framework originates from
Kauffman~\cite{Kauffman1986} and was subsequently studied
mathematically by Mossel and Steel~\cite{MosselSteel2005}.
Our model retains the same molecule and reaction structure,
but differs in its random assignment of catalysis.
In particular, Mossel and Steel assume independence of
catalysis events (R1) and reaction-independent catalysis
probabilities for each molecule (R2).

Hordijk, Wills and Steel~\cite{HordijkWillsSteel2014}
extended this framework to reaction-dependent catalysis
probabilities in the binary polymer model, with particular
attention to four specific catalysis mechanisms.
More recently, Hordijk and Steel~\cite{HordijkSteel2016}
investigated heterogeneous catalytic activity for arbitrary
finite alphabets, assigning to each molecule a random
number of reactions that it catalyses, with the reactions
selected uniformly.

Our model combines these two forms of heterogeneity
within a unified framework. Unlike the latter approach,
we assign random catalytic weights to molecules rather
than prescribing their numbers of catalysed reactions,
and allow reaction-dependent kernels. Conditional on the
weights, catalysis events are independent, whereas
unconditionally they may exhibit dependence through the
shared molecular weights. Moreover, the specific
formulation of the catalysis probabilities introduces
additional structure, making explicit that larger
catalytic weights and stronger molecule--reaction
compatibility lead to higher catalysis probabilities.
This enriches the model in terms of both biological
realism and mathematical tractability.


\subsection{Main results: RAF phase transitions under heterogeneous catalysis}
We next investigate under what conditions heterogeneous catalysis leads to the emergence of a RAF. Throughout this subsection, we consider the reaction system restricted to forward ligation reactions \(R_n^+\). Write 
\begin{align*}
   \mathbb P_n(\mathsf{RAF}):=\mathbb P(\mathsf{RAF}_n)
\end{align*}
for the probability that this system contains a RAF, where the probability is taken over both the molecular weights $W_x$'s and the catalysis indicators $I_{x,r}$'s.
Our aim is to identify conditions under which
\[
\mathbb P_n(\mathsf{RAF})\to 0 \qquad\text{or}\qquad\mathbb P_n(\mathsf{RAF})\to 1, \text{ as \(n\to\infty\). }
\]
As stated in our main theorem, Theorem \ref{thm:weighted-mass-criterion}, these conditions are expressed in terms of the total weighted catalytic mass available to each reaction and do not require a specific distribution for the molecular weights. 
For each \(r\in R_n^+\), define its {weighted catalytic mass} by 
\begin{align}\label{eq:molecular weights}
  M_n(r):=\sum_{x\in X_n}W_xK_n(x,r).  
\end{align}

Conditional on the sigma-algebra \(\mathcal W_n=\sigma(W_x:x\in X_n)\), the probability that \(r\) is uncatalysed is
\[
\begin{aligned} \mathbb P\bigl(I_{x,r}=0\text{ for every }x\in X_n \mid\mathcal W_n\bigr) = \prod_{x\in X_n} \exp\{-\lambda_nW_xK_n(x,r)\} = \exp\{-\lambda_nM_n(r)\}.
\end{aligned}
\]
Moreover, conditional on \(\mathcal W_n\), the events that distinct reactions are uncatalysed are mutually independent.
Define the set of forward reactions whose reactants belong to the food set by 
\begin{align}\label{eq:food set}
    \mathcal B_t := \{r\in R_n^+:\rho(r)\subseteq F_n\}
\end{align}

Since \(t\) is fixed, this set is independent of \(n\) for \(n\ge 2t\), with \(|\mathcal B_t|=|X_t|^2.\)
Every non-empty \(F_n\)-generated set of forward reactions contains at least one reaction from \(\mathcal B_t\). Thus, the catalytic masses of these reactions control an obstruction to RAF formation. Conversely, a sufficiently large catalytic mass for every forward reaction ensures RAF formation with probability tending to one. In our model, the analogues of Theorem~4.1 and Corollary~4.2 of Mossel and Steel~\cite{MosselSteel2005} read as follows. Throughout this paper, $\xrightarrow{\mathbb P}$ represents the convergence in probability.

\begin{theorem}[Weighted catalytic mass criterion]
\label{thm:weighted-mass-criterion}
Consider the weighted catalysis model on \(X_n\times R_n^+\), with arbitrary positive, finite, independent and identically distributed molecular weights. Let \(M_n(r)\) and \(\mathcal B_t\) be as in \eqref{eq:molecular weights} and \eqref{eq:food set}, respectively. Then we have the following phase transitions with two different regimes for the convergence of $\mathbb P_n(\mathsf{RAF})$.
\begin{enumerate}
\item \emph{Subcritical regime:}
If \(\lambda_n\max_{r\in\mathcal B_t}M_n(r) \xrightarrow{\mathbb P}0,\)
then \(\mathbb P_n(\mathsf{RAF})\to 0,\) as $n\to \infty$.
\item \emph{Supercritical regime:}
If \(\lambda_n\min_{r\in R_n^+}M_n(r) \xrightarrow{\mathbb P}\infty,\)
then \(\mathbb P_n(\mathsf{RAF})\to 1,\) as $n\to \infty$.
\end{enumerate}
\end{theorem}

The theorem separates the distributional properties of the weights from the reaction structure. That is, this theorem reduces the analysis of RAF formation to the asymptotic behaviour of the catalytic masses $\lambda_n M_n(r)$. To apply it, it suffices to estimate the weighted catalytic masses of food-generated reactions in the subcritical regime and those of all forward reactions in the supercritical regime.

Notably, the second part of Theorem \ref{thm:weighted-mass-criterion} allows for $\lambda_n \to 0$, showing that RAFs can emerge with high probability even when the baseline catalytic intensity vanishes. In this regime, a small number of molecules with exceptionally large catalytic weights may catalyze a substantial fraction of the reactions in an RAF.

\begin{figure}[!h]
    \centering
    \includegraphics[width=0.9\linewidth]{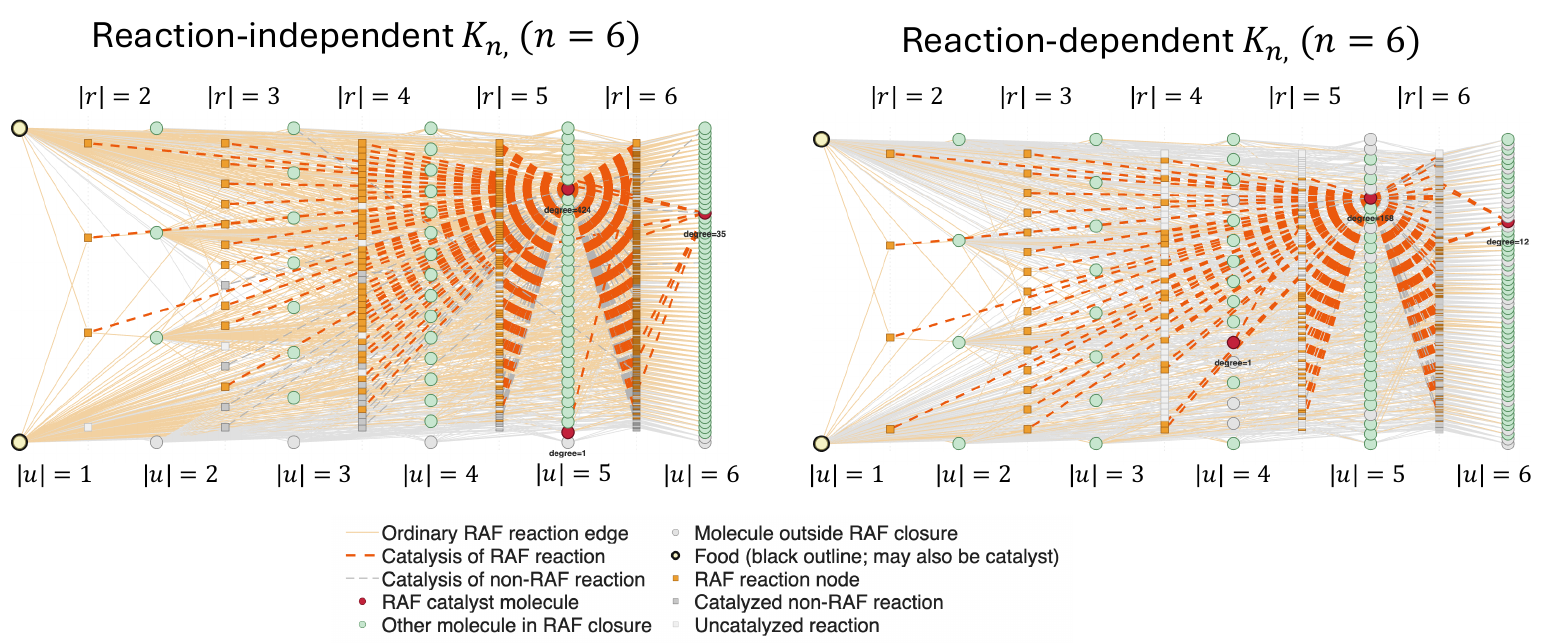}
    \caption{
    Examples of networks containing RAFs in the supercritical regime of Theorem \ref{thm:weighted-mass-criterion}, with $\lambda_n\to 0$ as $n\to\infty$. The left and right panels correspond to the reaction-independent and reaction-dependent kernels, respectively. For a reaction $r$ of the form $u+v\to uv$ or $uv\to u+v$, its size $|r|$ is defined as the length $|uv|$ of the polymer $uv$. Molecule and reaction nodes are vertically arranged according to their sizes. ``RAF catalyst molecules'' are molecules that catalyze at least one reaction in the RAF. The ``degree'' of a molecule denotes the number of reactions in the RAF that it catalyzes. 
    }
    \label{fig:raf samples}
\end{figure}

To illustrate this phenomenon, let
$\lambda_n=10^{-3}/\sqrt{\log |X_n|}$ and consider the reaction-independent kernel $K_n(r,x)=|X_n|$. We assume that the molecular weights $\{W_x\}_{x\in X_n}$ are independent and identically distributed Pareto random variables satisfying
\[
\mathbb P(W_x>w)=w^{-1/2}, \qquad w\ge 1.
\]
Under these assumptions, Theorem \ref{thm:weighted-mass-criterion} implies that $\mathbb P_n(\mathsf{RAF})\to 1$ as $n\to\infty$ (see Section \ref{sec:infinite mean} for details). Figure \ref{fig:raf samples} (left) shows a sample network generated under this setting that contains an RAF. Notably, only a few molecules catalyze a large proportion of the reactions in the RAF, reflecting the strong heterogeneity in molecular catalytic activity.

We next consider a reaction-dependent kernel while keeping the same $\lambda_n$ and distribution of molecular weights:
\[
K_n(r,x)=
\begin{cases}
0.1|X_n|, & \text{if } |r|\ge n/2,\\
|X_n|, & \text{if } |r|<n/2.
\end{cases}
\]
This setting also falls within the supercritical regime of Theorem \ref{thm:weighted-mass-criterion}. However, reactions with $|r|<n/2$ have greater catalytic affinity, making them more likely to participate in an RAF. This tendency is illustrated in Figure \ref{fig:raf samples} (right), where the RAF consists predominantly of reactions with $|r|<n/2$.

In the following sections, we explore interesting classes: the unweighted model and Pareto-distributed weights, whose phase transitions can be directly studied with Theorem \ref{thm:weighted-mass-criterion}.

\section{RAF Formation in Two Special Cases}

In this section, we introduce several classes of models for which phase transitions can be revealed by Theorem \ref{thm:weighted-mass-criterion}. These phase transitions are new compared to those of previous models. In particular, we provide an example where a single molecule can emerge RAF with asymptotic probability $1$ using specific $\{W_x\}_{x\in X_n}, \lambda_n$ and $K_n(x,r)$. 

\subsection{The unweighted model}\label{sec:un weighted}
The unweighted model arises from the general model by setting \(W_x\equiv 1\) for every \(x\in X_n\). The catalysis indicators are then mutually independent and satisfy
\[
I_{x,r}\sim\operatorname{Bernoulli} \left(1-\exp\{-\lambda_nK_n(x,r)\}\right), \qquad x\in X_n,\quad r\in R_n.
\]
Thus, heterogeneity in catalytic activity is determined entirely by the kernel \(K_n\). We next state sufficient conditions for the subcritical and supercritical regimes in this setting, in order to provide a baseline for comparison with the weighted model.

For this model, define the deterministic catalytic mass \(S_n(r):=\sum_{x\in X_n}K_n(x,r), \qquad r\in R_n^+.\)
The weighted catalytic mass criterion then applies with \(M_n(r)=S_n(r)\). Under uniform upper and lower bounds on these masses, the following proposition also provides explicit bounds on the RAF formation probability. As in Theorem~\ref{thm:weighted-mass-criterion}, we restrict attention to forward ligation reactions.
\begin{proposition}[RAF formation in the unweighted model]
\label{prop:unweighted-raf}
Consider the unweighted catalysis model on \(X_n\times R_n^+\), with food set \(F_n=X_t\), where \(t\ge 1\) is fixed. Write
\[
B_t:=|\mathcal B_t|, \qquad \mathcal B_t:=\{r\in R_n^+:\rho(r)\subseteq F_n\},
\]
so that \(B_t=|X_t|^2\) for \(n\ge 2t\). The following statements hold.
\begin{enumerate}
\item \emph{Subcritical bound:}
Suppose that $$\sup_n \sup _{r\in \mathcal{B}_t}  S_n(r)\le s_+.$$
Then, for every \(\lambda_n\ge 0\), \(\mathbb P_n(\mathsf{RAF}) \le 1-\exp\{-s_+B_t\lambda_n\}.\)
In particular, if \(\lambda_n\to 0\), then \(\mathbb P_n(\mathsf{RAF})\to 0\).
\item \emph{Supercritical bound:}
Suppose that $$\inf_n\inf_{r\in R_n^+}S_n(r)\geq s_->0.$$
If \(\lambda_n>(\log\kappa)/s_-\), then
\[
\mathbb P_n(\mathsf{RAF}) \ge 1- \frac{ \kappa\left(\kappa e^{-s_-\lambda_n}\right)^t }{ 1-\kappa e^{-s_-\lambda_n} }.
\]
Consequently, if \(\lambda_n\to\infty\), then \(\mathbb P_n(\mathsf{RAF})\to 1\). For a fixed intensity \(\lambda>(\log\kappa)/s_-\), the same lower bound holds uniformly in \(n>t\) and tends to \(1\) as \(\lambda\to\infty\).
\end{enumerate}
\end{proposition}
The subcritical bound requires control only of reactions whose reactants belong to the food set, whereas the supercritical bound uses a lower bound on the catalytic mass of every forward reaction. If both assumptions hold, vanishing intensity excludes RAF formation asymptotically, while diverging intensity ensures its occurrence with probability tending to one. Thus under this setting, small number of catalysts

The preceding result extends the analysis of RAF formation to
reaction-dependent catalysis probabilities. An important consequence
of allowing such dependence is that the classical criteria formulated
in terms of the expected number of reactions catalysed by each
molecule need no longer hold. 

To explain this distinction, recall Corollary~4.2 of
Mossel and Steel~\cite{MosselSteel2005}, which confirms the
conjectures of Steel~\cite{Steel2000}. Under their assumptions
(R1) and (R2), let $\mu_n(x)$ denote the expected number of
reactions catalysed by molecule $x$. They showed that
\begin{equation}
\label{eq:MS-subcritical}
\frac{1}{n}\max_{x\in X_n}\mu_n(x)
\longrightarrow 0
\quad\Longrightarrow\quad
 \mathbb P_n(\mathsf{RAF})\longrightarrow 0,
\end{equation}
whereas
\begin{equation}
\label{eq:MS-supercritical}
\min_{x\in X_n}
\frac{\mu_n(x)}{|x|}
\longrightarrow\infty
\quad\Longrightarrow\quad
\mathbb P_n(\mathsf{RAF})\longrightarrow 1.
\end{equation}
Here $|x|$ denotes the length of the molecule $x$.

A crucial assumption is (R2), which requires that, for each
fixed molecule $x$, the probability that $x$ catalyses a reaction
does not depend on the particular reaction. Consequently,
\[
\mu_n(x)
=
|R_n^+|\,\mathbb P((x,r)\in C_n),
\qquad r\in R_n^+.
\]
Thus, under (R2), the expected number of reactions catalysed
by a molecule also determines its probability of catalysing
any given reaction.

This correspondence is lost in the reaction-dependent kernel
model. Indeed, writing
\[
\mu_n(x)
=
\sum_{r\in R_n^+}
\left(1-e^{-\lambda_n K_n(x,r)}\right),
\]
the quantity $\mu_n(x)$ measures the total expected catalytic
activity of $x$, but does not specify how this activity is
distributed among reactions. The following proposition shows
that, without (R2), both implications
\eqref{eq:MS-subcritical} and \eqref{eq:MS-supercritical}
can fail, even when all catalysis indicators remain independent.

\begin{proposition}
\label{prop:failure-MS-criteria}
Suppose that the assumptions of Proposition \ref{prop:unweighted-raf} hold. Then neither of the implications
\eqref{eq:MS-subcritical} and \eqref{eq:MS-supercritical}
holds in general.

More precisely, the following two constructions are possible.

\begin{enumerate}
\item[(i)]
There exists a sequence of kernels such that
$
\dln \frac{1}{n}\max_{x\in X_n}\mu_n(x)
= 0,
$
while 
$
\dln \mathbb P_n(\mathsf{RAF})= 1.
$

\item[(ii)]
There exists a sequence of kernels such that
$
\dln \min_{x\in X_n}
\frac{\mu_n(x)}{|x|}
=\infty,
$
while 
$
\mathbb P_n(\mathsf{RAF})=0
$
for all sufficiently large $n$.
\end{enumerate}

In both cases, the kernels may be chosen with
$\lambda_n=1$ for every $n$.
\end{proposition}

The proposition demonstrates that reaction homogeneity of
the catalysis probabilities is essential to the classical
criteria. In the general kernel model, the total expected
catalytic activity of individual molecules does not determine
whether the reactions necessary for RAF formation receive
sufficient catalytic support. This motivates the use of
reaction-wise catalytic mass in the preceding theorem.

\subsection{Heavy-tailed catalytic activities}
We now assume that the molecular weights have a regularly varying tail. This assumption allows for substantial heterogeneity in catalytic activity, with a small proportion of exceptionally active molecules whose catalytic propensities far exceed those of typical molecules. Compared with the homogeneous random catalysis model studied by Mossel and Steel \cite{MosselSteel2005}, our model incorporates heterogeneous, molecule-dependent catalytic propensities. This extension is analogous to the transition from the classical Erd"os--R'enyi random graph model to the Norros--Reittu model \cite{norros2006conditionally}, in which vertex-specific weights determine connection probabilities and allow for highly heterogeneous degree distributions.

\begin{assumption}[Regularly varying catalytic weights]
\label{ass:heavy-tailed-weights}
There exist \(\alpha>0\) and a measurable, slowly varying function at infinty \(L:(0,\infty)\to(0,\infty)\) such that
\begin{equation}
\lim_{w\to\infty}
\frac{\mathbb P(W>w)}{w^{-\alpha}L(w)}=1.
\label{eq:regular-variation}
\end{equation}
\end{assumption}
Recall that \(L\) is slowly varying at infinity if $\frac{L(cw)}{L(w)}\rightarrow 1$, as $w\to\infty$
for every \(c>0\). The index \(\alpha\) determines the moment behaviour of \(W\): for every \(p>0\),
\[
p<\alpha\quad\Longrightarrow\quad\mathbb E[W^p]<\infty, \qquad p>\alpha\quad\Longrightarrow\quad\mathbb E[W^p]=\infty.
\]
Consequently, \(0<\alpha<1\) gives an infinite mean, \(1<\alpha<2\) gives a finite mean and infinite variance, and \(\alpha>2\) gives finite first and second moments. At \(p=\alpha\), moment finiteness depends on the slowly varying factor. In particular, the cases \(\alpha=1\) and \(\alpha=2\) require separate consideration when determining the mean and variance.
Molecule-specific weights allow catalytic propensities to vary even among molecules with the same compatibility kernel. Under Assumption~\ref{ass:heavy-tailed-weights}, some molecules may therefore have substantially greater catalytic activity than a typical molecule. We next examine how this heterogeneity is reflected in the number of reactions catalysed.
Consider a reaction-independent compatibility kernel, \(K_n(x,r)=k_n(x), \qquad x\in X_n,\quad r\in R_n^+,\)
and fix \(\lambda_n\equiv\lambda>0\). Write \(r_n:=|R_n^+|\) and define \(D_n(x):=\sum_{r\in R_n^+}I_{x,r},\)
the number of forward reactions catalysed by molecule \(x\). Conditional independence of the catalysis indicators implies
\[
D_n(x)\mid W_x \sim \operatorname{Bin}\left( r_n,\, 1-e^{-\lambda W_xk_n(x)} \right).
\]
In particular,
\[
\mathbb E[D_n(x)\mid W_x] = r_n\left(1-e^{-\lambda W_xk_n(x)}\right) \qquad\text{almost surely}.
\]
For fixed \(n\), the count \(D_n(x)\) is bounded by \(r_n\), so its distribution cannot have a regularly varying tail at infinity. The appropriate statement concerns growing thresholds below this finite-size bound. With the normalization \(k_n(x)=r_n^{-1}\), the following theorem shows that the catalytic count inherits the tail index of the molecular weight.
\begin{theorem}[Power-law for the number of catalysed reactions]
\label{thm:catalytic-count-tail}
Suppose that Assumption~\ref{ass:heavy-tailed-weights} holds, that \(\lambda_n\equiv\lambda>0\), and that $K_n(x,r)=\frac{1}{r_n}$, for each $x\in X_n$ and for each $r\in R_n^+,$
where \(r_n:=|R_n^+| =\sum_{\ell=2}^{n}(\ell-1)\kappa^\ell.\)
Let \(x_n\in X_n\) be any sequence of molecules. For every sequence of positive real numbers \(t_n\) satisfying $t_n\to \infty$ and $\displaystyle \frac{t_n}{r_n}\to 0$, as $n\to \infty$,
we have that
\begin{equation}
\label{eq:catalytic-count-tail} \mathbb P(D_n(x_n)>t_n) \sim \overline F\left(\frac{t_n}{\lambda}\right) \sim \lambda^\alpha t_n^{-\alpha}L(t_n), \qquad n\to\infty,
\end{equation}
where $\overline F(w):=P(W>w)$. Moreover, for every fixed \(s>0\),
\begin{equation}
\label{eq:catalytic-count-tail-ratio} \frac{\mathbb P(D_n(x_n)>s t_n)} {\mathbb P(D_n(x_n)>t_n)} \longrightarrow s^{-\alpha}.
\end{equation}
\end{theorem}
Thus, at thresholds tending to infinity but remaining small relative to \(r_n\), the number of reactions catalysed by a molecule exhibits the same regularly varying tail index as its weight. The normalization \(K_n(x,r)=r_n^{-1}\) makes this correspondence transparent: whenever \(W_x/r_n\) is small, the conditional mean of \(D_n(x)\) is asymptotic to \(\lambda W_x\).

\subsection{RAF formation with Pareto weights}
We now examine how random molecular weights change the intensity scale for RAF formation relative to the unweighted model. Under the kernel assumptions stated below, finite-mean weights lead to the same asymptotic intensity scale, up to a constant factor. Infinite-mean weights, by contrast, can produce sufficiently large catalytic masses for RAFs to emerge with probability tending to one even when the global intensity \(\lambda_n\) tends to zero.

The weighted-mass criterion in Theorem~\ref{thm:weighted-mass-criterion} reduces the analysis of RAF formation to the asymptotic behaviour of the catalytic masses $\lambda_n M_n(r)$. Consequently, different distributions of the molecular weights can be studied within the same general framework,
without requiring separate RAF formation arguments. The remaining task is therefore to determine the asymptotic scale of \(M_n(r)=\sum_{x\in X_n}W_xK_n(x,r).\)

We specialize to Pareto-distributed weights satisfying \(\mathbb P(W>w)=w^{-\alpha}, w\ge 1,\)
where \(\alpha>0\). We distinguish three regimes: the finite-mean case \(\alpha>1\), the critical infinite-mean case \(\alpha=1\), and the infinite-mean case \(0<\alpha<1\). Their different weighted-sum asymptotics yield corresponding criteria for the absence and emergence of RAFs.

Let us remark that these criteria can also be recovered from Theorem~1 of Hordijk and Steel~\cite{HordijkSteel2016} by expressing the expected number of catalysed reactions per
molecule in terms of the Laplace transform of the Pareto distribution. We derive them directly from
our weighted-mass criterion in Theorem~\ref{thm:weighted-mass-criterion}, which also accommodates
reaction-dependent kernels.

\subsubsection{The finite-mean regime: \texorpdfstring{\(\alpha>1\)}{alpha > 1}}

Suppose first that \(\alpha>1\). Then \(\mathbb E[W] = \frac{\alpha}{\alpha-1}.\)
For a reaction \(r\), write \(S_n(r) = \sum_{x\in X_n}K_n(x,r).\) Under this regime, we have the same phase transition as the unweighted model shown in Section \ref{sec:un weighted}.

\begin{proposition}[Finite-mean regime phase transition]\label{prop:finitemean}
	Suppose that \(\alpha>1\). Then
    \begin{enumerate}
        \item if \(\sup_n\sup_{r\in\mathcal B_t}S_n(r)<\infty\)
	and \(\dln \lambda_n= 0,\)
	then \( \dln \mathbb P_n(\mathrm{RAF})\to 0\), and

    \item if \(\inf_n\inf_{r\in R_n^+}S_n(r)\geq s_->0\)
	and \( 
    \dln \lambda_n= \infty,\)
	then \( \dln \mathbb P_n(\mathrm{RAF})\to 1\).
    \end{enumerate}

\end{proposition}

\subsubsection{The critical infinite-mean regime: \texorpdfstring{\(\alpha=1\)}{alpha = 1}}

We next consider the critical Pareto case $\mathbb P(W>w)=w^{-1}$ with $w\geq1$.
For the sake of simplicity, we suppose in this subsection that the kernel is
reaction-independent: \(K_n(x,r)=k_n(x).\)
Then every reaction has the same weighted catalytic mass, \(M_n(r)=M_n := \sum_{x\in X_n}k_n(x)W_x.\)
Recall that \(S_n :=\sum_{x\in X_n}K_n(x,r)= \sum_{x\in X_n}k_n(x)\)
and define \(J_n(\lambda_n) := \sum_{x\in X_n} k_n(x) \log\frac{1}{\lambda_nk_n(x)},\)
with the convention that the summand is zero whenever \(k_n(x)=0\). Then we can set a phase transition parameter $A^{(1)}_n := \lambda_n J_n(\lambda_n)$. 

\begin{theorem}[Critical Pareto phase transition]\label{thn:infcrit}
	Suppose that \(\alpha=1\), \(K_n(x,r)=k_n(x),\)
	and \(\sup_n S_n<\infty.\)
	Assume also that \(\lambda_n\to 0,\) as $n\to \infty$.
	Then 
    \begin{enumerate}
        \item \(\dln A^{(1)}_n =0\)
	implies that \( \dln \mathbb P_n(\mathrm{RAF})=0,\) and
    \item \( \dln A^{(1)}_n  =\infty\)
	implies that \( \dln \mathbb P_n(\mathrm{RAF})=1\).
    \end{enumerate}
\end{theorem}

\subsubsection{The infinite-mean regime: \texorpdfstring{\(0<\alpha<1\)}{0 < alpha < 1}}\label{sec:infinite mean}

Suppose now that \(0<\alpha<1\)
and again assume a reaction-independent kernel \(K_n(x,r)=k_n(x).\) Then $A_n^{(\alpha)}
:=
\lambda_n^\alpha
\sum_xk_n(x)^\alpha$ provides a phase transition parameter. 

\begin{theorem}[Infinite-mean Pareto phase transition]\label{thm:inf}
	Suppose that \(0<\alpha<1\) and \(K_n(x,r)=k_n(x).\) Then
    \begin{enumerate}
        \item \(\dln A_n^{(\alpha)} =0\)
	implies that \( \dln \mathbb P_n(\mathrm{RAF})=0,\) and
    \item \( \dln A_n^{(\alpha)}  =\infty\)
	implies that \( \dln \mathbb P_n(\mathrm{RAF})=1\).
    \end{enumerate}    
\end{theorem}

\subsubsection{Example: a diffuse reaction-independent kernel}\label{subsubsec334}

The three regimes can be compared particularly clearly with a uniform catalysis kernel $K_n(x,r)=\frac1{N_n}$, where $N_n=|X_n|$. 

Since $S_n(r)=1$ for any $r$ in this case, when \(\alpha>1\), $\lambda_n$ solely determines the phase transition parameter by Proposition \ref{prop:finitemean}.
More generally, in Section \ref{sec:finitemean proof} we can see that for any fixed \(\lambda>\log\kappa,\)
the finite-mean argument gives
\[
\mathbb P_n(\mathsf{RAF}) \geq 1- \frac{ \kappa(\kappa e^{-\lambda})^t }{ 1-\kappa e^{-\lambda} }.
\]
For \(\alpha=1\), the phase parameter is \(\lambda_n \log\frac{N_n}{\lambda_n}\) by Theorem \ref{thn:infcrit}. An interesting case comes with the vanishing intensity, $\dln \lambda_n=0$. In this case, \(\dln \lambda_n\log\frac1{\lambda_n}= 0,\)
and hence $\lambda_n\log N_n$ determines the phase transition due to $
\lambda_n \log\frac{N_n}{\lambda_n} = \lambda_n\log N_n + \lambda_n\log\frac1{\lambda_n}$. 
For \(0<\alpha<1\), \(\sum_{x\in X_n}k_n(x)^\alpha=\sum_x \left(\frac1{N_n}\right)^\alpha = N_n^{1-\alpha},\)
so the phase parameter is \(A_n^{(\alpha)} = \lambda_n^\alpha N_n^{1-\alpha}\) by Theorem \ref{thm:inf}.

The diffuse model therefore exhibits the following three characteristic
scales:
\[
\begin{array}{c|c} \text{weight regime} & \text{phase parameter} \\ \hline \alpha>1 & \lambda_n \\[1mm] \alpha=1 & \lambda_n\log N_n \text{ when $\lambda_n\to 0$, as $n\to \infty$}\\[1mm] 0<\alpha<1 & \lambda_n^\alpha N_n^{1-\alpha}.
\end{array}
\]

This comparison shows that the distribution of catalytic activity can
change the asymptotic RAF threshold itself. In particular, a vanishing intensity does not necessarily suppress RAF formation
in the infinite-mean regimes. For example, in the critical case \(\alpha=1\), take \(\lambda_n = \frac{1}{\sqrt{\log N_n}}.\)
Then \(\dln \lambda_n= 0,\)
but \(\dln \lambda_n\log N_n = \sqrt{\log N_n} =\infty.\)
Consequently, \(\mathbb \dln P_n(\mathsf{RAF})=1.\)
The same sequence \(\lambda_n\to0\) is subcritical in the finite-mean
regime \(\alpha>1\).
Similarly, if \(0<\alpha<1\), choose \(\lambda_n=N_n^{-\beta}\)
with \(0<\beta<\frac{1-\alpha}{\alpha}.\)
Then \(\dln \lambda_n= 0,\)
while
\[
\dln \lambda_n^\alpha N_n^{1-\alpha} = \dln N_n^{1-\alpha-\alpha\beta} =\infty.
\]
Hence once again \(\dln \mathbb P_n(\mathsf{RAF})= 1\)
despite the fact that the global catalytic intensity tends to zero.

Thus a sequence of global catalytic intensities that is asymptotically
subcritical in the finite-mean regime can be asymptotically
supercritical when the distribution of catalytic activity has infinite
mean. The heavy-tailed distribution of catalytic potential therefore
does more than increase the probability of RAF formation: it changes
the scale of the RAF phase transition.

\begin{remark}[Reaction-dependent catalysis]
The preceding phase-transition results extend to
reaction-dependent kernels. Indeed, for a forward ligation reaction $r:a+b\to ab$, consider
the kernel
\[
K_n(x,r)
=
\frac{1+\mathbf 1\{x_{|x|}=(ab)_1\}}{N_n},
\qquad x\in X_n,
\]
where $x_{|x|}$ denotes the last symbol of $x$,
$(ab)_1$ denotes the first symbol of the product,
and $N_n=|X_n|$.
Thus, catalysis is more likely when the last symbol
of the candidate catalyst agrees with the first
symbol of the reaction product.

Writing
\[
\overline W_n
:=
\frac{1}{N_n}\sum_{x\in X_n}W_x,
\]
the corresponding weighted catalytic masses satisfy
\[
\overline W_n
\leq M_n(r)
=\sum_{x\in X_n}W_xK_n(x,r)
\leq 2\overline W_n
\]
uniformly over all reactions $r\in R_n^+$.
Consequently, Theorem~\ref{thm:weighted-mass-criterion}
implies the same phase-transition criteria as for the
reaction-independent diffuse kernel:
\[
\begin{array}{ll}
\alpha>1:
&
\begin{cases}
\lambda_n\to0
&\Longrightarrow \mathbb P_n(\mathsf{RAF})\to0, \text{ and} \\
\lambda_n\to\infty
&\Longrightarrow \mathbb P_n(\mathsf{RAF})\to1, \text{ as $n\to \infty$};
\end{cases}
\\[3ex]
\alpha=1:
&
\begin{cases}
\lambda_n\log N_n\to0
&\Longrightarrow \mathbb P_n(\mathsf{RAF})\to0 \text{ and},\\
\lambda_n\log N_n\to\infty
&\Longrightarrow \mathbb P_n(\mathsf{RAF})\to1 \text{ as $n\to \infty$};
\end{cases}
\\[3ex]
0<\alpha<1:
&
\begin{cases}
\lambda_n^\alpha N_n^{1-\alpha}\to0
&\Longrightarrow \mathbb P_n(\mathsf{RAF})\to0 \text{ and},\\
\lambda_n^\alpha N_n^{1-\alpha}\to\infty
&\Longrightarrow \mathbb P_n(\mathsf{RAF})\to1, \text{ as $n\to \infty$}.
\end{cases}
\end{array}
\]

This example shows that the phase-transition regimes
obtained for the diffuse kernel remain valid in the
presence of nontrivial molecule--reaction compatibility.
More generally, the same conclusions hold for any
kernel satisfying
\[
\frac{c}{N_n}\leq K_n(x,r)\leq\frac{C}{N_n}
\]
uniformly in $x$, $r$, and $n$, for constants
$0<c\leq C<\infty$.
\end{remark}

We conclude this section with an example illustrating a
particularly striking consequence of heavy-tailed catalytic
potential. In the infinite-mean regime, RAF formation can be
driven by a single exceptionally active molecule. More
precisely, we show that, under a suitable scaling of the
catalytic intensity, one molecule can catalyse every reaction
of an $F$-generated RAF with probability tending to one,
even when the global catalytic intensity vanishes.

\begin{proposition}
\label{prop:single-catalyst-raf}
Let the assumptions in Section~\ref{subsubsec334} hold. Suppose that
\[ \dln
\frac{\lambda_n^\alpha N_n^{1-\alpha}}
{(\log N_n)^\alpha}
=\infty.
\]
Then, with probability tending to one, there exists
a nonempty $F$-generated RAF whose every reaction
is catalysed by the same molecule for arbitrary $\alpha$.
In particular,
\[
\dln \mathbb P_n(\mathsf{RAF})= 1.
\]
\end{proposition}

\section{Proofs}

In this section, we first prove Theorem~\ref{thm:catalytic-count-tail}. We then establish Proposition~\ref{prop:unweighted-raf}, where the underlying RAF argument is most transparent, before extending the argument to prove Theorem~\ref{thm:weighted-mass-criterion}. Finally, we prove the results for Pareto-distributed weights, Proposition~\ref{prop:finitemean} and Theorems~\ref{thn:infcrit} and~\ref{thm:inf}.

\subsection{Proof of Theorem~\ref{thm:catalytic-count-tail}}

\begin{proof}
	Conditional on \(W_x=w\), \(D_n(x) \sim \operatorname{Bin}\left( r_n,1-e^{-\lambda w/r_n} \right).\) Let \(Y_{n,w} \overset{d}{=} \bigl(D_n(x)\mid W_x=w\bigr).\) Its conditional mean is \(\mu_n(w) := \mathbb E[Y_{n,w}] = r_n\left(1-e^{-\lambda w/r_n}\right),\) and its conditional variance is

	$$
\operatorname{Var}(Y_{n,w}) = r_n \left(1-e^{-\lambda w/r_n}\right) e^{-\lambda w/r_n}.
$$

	In particular, \(\operatorname{Var}(Y_{n,w}) \le \mu_n(w).\) For \(t>1\), set \(\delta_t:=\sqrt{t}\log t\) and define \(m_n(t)\) by \(\mu_n(m_n(t))=t-\delta_t.\) Solving explicitly gives

	$$
m_n(t) = -\frac{r_n}{\lambda} \log\left( 1-\frac{t-\sqrt{t}\log t}{r_n} \right).
$$

	Since \(t_n/r_n\to0\), \(m_n(t_n) \sim \frac{t_n}{\lambda}.\) More generally, for every fixed \(s>0\), \(\frac{m_n(s t_n)}{m_n(t_n)} \longrightarrow s.\) By conditioning on \(W_x\),
\[
\mathbb P(D_n(x)>t) = \int_0^\infty \mathbb P(Y_{n,w}>t)\,dF(w).
\]

	Split the integral into

	$$
I_1=[0,m_n(t)) \qquad\text{and}\qquad I_2=[m_n(t),\infty).
$$

	For \(w\in I_1\), \(\mu_n(w) \le t-\sqrt{t}\log t.\) Bernstein's inequality therefore gives

	$$
\mathbb P(Y_{n,w}>t) \le \exp\left( -\frac{(t-\mu_n(w))^2} {2\mu_n(w)+4t/3} \right) \le \exp\left( -\frac{3}{10}(\log t)^2 \right).
$$

	Hence, for every \(a>0\), 
\[
t^a \int_{I_1} \mathbb P(Y_{n,w}>t)\,dF(w) \longrightarrow 0.
\]

	Define 
\[
G_n(t) := \int_{w>m_n(t)} \mathbb P(Y_{n,w}>t)\,dF(w).
\]

	Then \(\mathbb P(D_n(x)>t) = G_n(t)+E_n(t),\) where \(E_n(t)=o(t^{-a})\)
	for every \(a>0\). Clearly, \(G_n(t) \le \overline F(m_n(t)).\) For the converse estimate, fix \(\varepsilon>0\). Then
\[
G_n(t) \ge \int_{w>(1+\varepsilon)m_n(t)} \mathbb P(Y_{n,w}>t)\,dF(w),
\]

	and consequently

	$$
G_n(t) \ge \overline F((1+\varepsilon)m_n(t)) - \int_{w>(1+\varepsilon)m_n(t)} \mathbb P(Y_{n,w}\le t)\,dF(w).
$$

	We next estimate the last integral. Since \(\mu_n(w)\) is increasing in \(w\), for \(w>(1+\varepsilon)m_n(t)\) we have \(\mu_n(w) \ge \mu_n((1+\varepsilon)m_n(t)).\) By the definition of \(m_n(t)\), \(e^{-\lambda m_n(t)/r_n} = 1-\frac{t-\delta_t}{r_n}.\) Thus

	$$
\mu_n((1+\varepsilon)m_n(t)) = r_n \left[ 1- \left( 1-\frac{t-\delta_t}{r_n} \right)^{1+\varepsilon} \right].
$$

	Since \(t/r_n\to0\), \(\mu_n((1+\varepsilon)m_n(t)) = (1+\varepsilon)t+o(t).\) It follows that, for all sufficiently large \(n\), \(\mu_n(w) \ge \left(1+\frac{\varepsilon}{2}\right)t\) uniformly for \(w>(1+\varepsilon)m_n(t).\) By Chebyshev's inequality,

	$$
\mathbb P(Y_{n,w}\le t) \le \frac{\operatorname{Var}(Y_{n,w})} {(\mu_n(w)-t)^2} \le \frac{\mu_n(w)} {(\mu_n(w)-t)^2}.
$$

	For \(u>t\), the function \(u\longmapsto \frac{u}{(u-t)^2}\) is decreasing. Hence there exists a constant \(C_\varepsilon<\infty\) such that \(\mathbb P(Y_{n,w}\le t) \le \frac{C_\varepsilon}{t}\) uniformly for \(w>(1+\varepsilon)m_n(t).\) Therefore

	$$
\int_{w>(1+\varepsilon)m_n(t)} \mathbb P(Y_{n,w}\le t)\,dF(w) \le \frac{C_\varepsilon}{t} \overline F((1+\varepsilon)m_n(t)).
$$

	Consequently, \(G_n(t) \ge \left(1-o(1)\right) \overline F((1+\varepsilon)m_n(t)).\) Combining the upper and lower estimates gives

	$$
\left(1-o(1)\right) \overline F((1+\varepsilon)m_n(t)) \le G_n(t) \le \overline F(m_n(t)).
$$

	Since \(\overline F\) is regularly varying with index \(-\eta\),

	$$
\frac{ \overline F((1+\varepsilon)m_n(t)) }{ \overline F(m_n(t)) } \longrightarrow (1+\varepsilon)^{-\eta}.
$$

	Letting first \(n\to\infty\) along the sequence \(t=t_n\), and then \(\varepsilon\downarrow0\), yields \(G_n(t_n) \sim \overline F(m_n(t_n)).\) Moreover, \(m_n(t_n)\sim\frac{t_n}{\lambda},\) so regular variation gives
\[
\overline F(m_n(t_n)) \sim \overline F\left(\frac{t_n}{\lambda}\right).
\]

	The contribution from \(I_1\) is negligible compared with this quantity. Indeed, it decays faster than every power of \(t_n\), whereas \(\overline F\left(\frac{t_n}{\lambda}\right)\) is regularly varying with index \(-\eta\). Therefore
\[
\mathbb P(D_n(x)>t_n) \sim \overline F\left(\frac{t_n}{\lambda}\right).
\]

	Finally,

	$$
\overline F\left(\frac{t_n}{\lambda}\right) = \left(\frac{t_n}{\lambda}\right)^{-\eta} L\left(\frac{t_n}{\lambda}\right) \sim \lambda^\eta t_n^{-\eta}L(t_n),
$$

	since \(L\) is slowly varying. For fixed \(s>0\),

	$$
\frac{\mathbb P(D_n(x)>s t_n)} {\mathbb P(D_n(x)>t_n)} \sim \frac{ \overline F(s t_n/\lambda) }{ \overline F(t_n/\lambda) } \longrightarrow s^{-\eta}.
$$

	This proves the result.
\end{proof}

\subsection{Proof of Proposition~\ref{prop:unweighted-raf}}

\begin{lemma}\label{lemm:unweighted}
Fix \(n\in\mathbb N\). Let \(X_n\) be the molecular set and \(R_n\) the reaction set, and let \(K_n:X_n\times R_n\to[0,\infty)\)

be a deterministic catalytic compatibility kernel. For \(\lambda_n\ge 0\), suppose that the catalysis indicators \(\{I_{x,r}:x\in X_n,\ r\in R_n\}\)

are mutually independent and satisfy \(\mathbb P(I_{x,r}=1) = 1-\exp\{-\lambda_nK_n(x,r)\}.\)

For each reaction \(r\in R_n\), define \(S_n(r) := \sum_{x\in X_n}K_n(x,r)\)

and let

$$
q_n(r) := \mathbb P\bigl( I_{x,r}=0 \text{ for every }x\in X_n \bigr)
$$
denote the probability that \(r\) receives no catalyst. Then:

\begin{enumerate}

\item[(i)] For every \(r\in R_n\), $q_n(r) = \exp\{-\lambda_nS_n(r)\}$.

\item[(ii)] Suppose there exist constants
\(0<s_-\le s_+<\infty\)
such that, for every sufficiently large \(n\) and every \(r\in R_n\), \(s_- \le S_n(r) \le s_+.\)
Then, uniformly in \(r\in R_n\), \(e^{-s_+\lambda_n} \le q_n(r) \le e^{-s_-\lambda_n}.\)

Consequently, if \(\lambda_n\to0\), then \(\inf_{r\in R_n}q_n(r)\to1,\)
whereas if \(\lambda_n\to\infty\), then \(\sup_{r\in R_n}q_n(r)\to0.\)

\item[(iii)] More generally, let \(a_n,b_n\ge0\). Under the previous assumptions, if

\(\lambda_n\le a_n,\)
then
$$
q_n(r)\ge e^{-s_+a_n} \qquad \text{for every }r\in R_n,
$$

while if \(\lambda_n\ge b_n,\)
then

$$
q_n(r)\le e^{-s_-b_n} \qquad \text{for every }r\in R_n.
$$

\end{enumerate}
\end{lemma}

\begin{proof}
Fix \(r\in R_n\). For each \(x\in X_n\),
\[
\mathbb P(I_{x,r}=0) = 1-\mathbb P(I_{x,r}=1) = \exp\{-\lambda_nK_n(x,r)\}.
\]

Since the family \(\{I_{x,r}:x\in X_n\}\) is independent, the probability that no molecule catalyses \(r\) is the product of the corresponding failure probabilities. Hence \(q_n(r) = \prod_{x\in X_n} \mathbb P(I_{x,r}=0).\) Therefore,
\[
q_n(r) = \prod_{x\in X_n} \exp\{-\lambda_nK_n(x,r)\}.
\]

Using the identity \(\prod_{x\in X_n}e^{-a_x} = e^{-\sum_{x\in X_n}a_x},\) we obtain
\[
q_n(r) = \exp\left\{ -\lambda_n \sum_{x\in X_n}K_n(x,r) \right\}.
\]

By the definition of \(S_n(r)\), \(q_n(r) = \exp\{-\lambda_nS_n(r)\}.\) This proves part \textup{(i)}. For part \textup{(ii)}, assume \(s_-\le S_n(r)\le s_+.\) Since \(\lambda_n\ge0\), \(-\lambda_ns_+ \le -\lambda_nS_n(r) \le -\lambda_ns_-.\) The exponential function is increasing, and therefore \(e^{-s_+\lambda_n} \le e^{-\lambda_nS_n(r)} \le e^{-s_-\lambda_n}.\) Using part \textup{(i)}, \(e^{-s_+\lambda_n} \le q_n(r) \le e^{-s_-\lambda_n}.\) The bounds are uniform in \(r\), hence \(\inf_{r\in R_n}q_n(r) \ge e^{-s_+\lambda_n}.\) If \(\lambda_n\to0\), then \(e^{-s_+\lambda_n}\to1.\) Since always \(q_n(r)\le1\), it follows that \(\inf_{r\in R_n}q_n(r)\to1.\) Similarly, \(\sup_{r\in R_n}q_n(r) \le e^{-s_-\lambda_n}.\) If \(\lambda_n\to\infty\), then, because \(s_->0\), \(e^{-s_-\lambda_n}\to0,\)
and therefore \(\sup_{r\in R_n}q_n(r)\to0.\) This proves part \textup{(ii)}. For part \textup{(iii)}, suppose first that \(\lambda_n\le a_n.\) From part \textup{(ii)}, \(q_n(r)\ge e^{-s_+\lambda_n}.\) Since the function \(t\mapsto e^{-s_+t}\) is decreasing, \(e^{-s_+\lambda_n} \ge e^{-s_+a_n},\)
and hence \(q_n(r)\ge e^{-s_+a_n}.\) Likewise, if \(\lambda_n\ge b_n,\)
then \(q_n(r)\le e^{-s_-\lambda_n} \le e^{-s_-b_n}.\) This proves part \textup{(iii)}.
\end{proof}

\begin{corollary}[Normalized kernel]
Suppose that

$$
\sum_{x\in X_n}K_n(x,r)=1 \qquad \text{for every }n \text{ and every }r\in R_n.
$$

Then \(q_n(r)=e^{-\lambda_n}\)
for every \(r\in R_n\). In particular, \(\lambda_n\to0 \quad\Longrightarrow\quad q_n(r)\to1\)

uniformly in \(r\), while \(\lambda_n\to\infty \quad\Longrightarrow\quad q_n(r)\to0\)

uniformly in \(r\).
\end{corollary}

\begin{proof}
Under the normalization assumption, \(S_n(r)=1\)
for every \(r\in R_n\). Substituting into part \textup{(i)} of the preceding lemma gives \(q_n(r) = \exp\{-\lambda_n\} = e^{-\lambda_n}.\) The two limiting statements follow immediately.
\end{proof}

\begin{proof}[Proof of Proposition~\ref{prop:unweighted-raf}]
    We first prove the subcritical bound. For \(r\in R_n^+\), let
\[
q_n(r) := \mathbb{P}\left( I_{x,r}=0 \text{ for every } x\in X_n \right)
\]
denote the probability that \(r\) is not catalysed by any molecule in \(X_n\). By mutual independence of the catalysis indicators and the definition of the model,
\[
\begin{aligned} q_n(r) &= \prod_{x\in X_n} \mathbb{P}(I_{x,r}=0)\\ &= \prod_{x\in X_n} \exp\{-\lambda_n K_n(x,r)\}\\ &= \exp\left\{ -\lambda_n \sum_{x\in X_n} K_n(x,r) \right\}.
\end{aligned}
\]
By assumption and Lemma~\ref{lemm:unweighted},
\begin{equation}
q_n(r)\ge \exp\{-s_+\lambda_n\} \qquad \text{for every } r\in R_n^+. \tag{1}
\end{equation}

We next establish a necessary condition for the existence of a RAF. Suppose that \(R'\subseteq R_n^+\) is a RAF. Since \(R'\) is nonempty and \(F\)-generated, every reactant of every reaction in \(R'\) belongs to
\(\operatorname{cl}_{R'}(F).\)
By definition, \(\operatorname{cl}_{R'}(F)\) is obtained from \(F\) by
successively adjoining the products of reactions in \(R'\) whose reactants
are already present. Because \(R'\) is nonempty and \(F\)-generated, at least one reaction \(r\in R'\) must be applicable at the first stage of this construction.
Thus \(\rho(r)\subseteq F,\)
so that \(r\in\mathcal{B}_t.\) Since \(R'\) is reflexively autocatalytic, \(r\) is catalysed by at least one
molecule in \(\operatorname{cl}_{R'}(F)\subseteq X_n.\)
Hence \(r\) is globally catalysed. Therefore,
\begin{equation}
\{\text{a RAF exists}\} \subseteq \left\{ \text{at least one reaction in }\mathcal{B}_t \text{ is globally catalysed} \right\}. \tag{2}
\end{equation}
Equivalently,
\begin{equation}
\left\{ \text{every reaction in }\mathcal{B}_t \text{ is globally uncatalysed} \right\} \subseteq \{\text{no RAF exists}\}. \tag{3}
\end{equation}

For distinct reactions \(r,r'\in\mathcal{B}_t\), the events
\[
\{r\text{ is globally uncatalysed}\} \qquad\text{and}\qquad \{r'\text{ is globally uncatalysed}\}
\]
are independent. Indeed, the first event depends only on \(\{I_{x,r}:x\in X_n\},\)
whereas the second depends only on \(\{I_{x,r'}:x\in X_n\}.\)
These are disjoint subfamilies of the mutually independent collection \(\{I_{x,r}:x\in X_n,\ r\in R_n^+\}.\) It follows from \textup{(3)} that
\[
\begin{aligned} 1-\mathbb P_n(\mathsf{RAF}) &\ge \mathbb{P}\left( \text{every } r\in\mathcal{B}_t \text{ is globally uncatalysed} \right)\\ &= \prod_{r\in\mathcal{B}_t} q_n(r).
\end{aligned}
\]
Applying \textup{(1)},
\[
1-\mathbb P_n(\mathsf{RAF}) \ge \prod_{r\in\mathcal{B}_t} e^{-s_+\lambda_n} = e^{-s_+B_t\lambda_n}.
\]
Consequently, \(\mathbb P_n(\mathsf{RAF}) \le 1-e^{-s_+B_t\lambda_n}.\) Finally, \(s_+\) and \(B_t\) are independent of \(n\) for all sufficiently
large \(n\). Hence, if \(\lambda_n\to 0,\)
then \(s_+B_t\lambda_n\to 0,\)
and therefore \(1-e^{-s_+B_t\lambda_n}\to 0.\)
Thus \(\mathbb P_n(\mathsf{RAF})\to 0.\) It remains to prove the supercritical bound. By the assumed lower bound and Lemma~\ref{lemm:unweighted}, \(q_n(r) \le e^{-s_-\lambda}\)
	for every \(r\in R_n^+\). Set \(q=e^{-s_-\lambda}.\) The condition \(\lambda>\frac{\log\kappa}{s_-}\)
	is equivalent to \(\kappa q<1.\) We now consider the recursive construction of the molecules in \(X_n\) from the food set \(F=X(t)\). Fix \(s\in\{t,\ldots,n-1\}\), and suppose that all molecules in \(X(s)\) have already been generated. Let \(x=x_1x_2\cdots x_{s+1}\in\mathcal A^{s+1}\) be a molecule of length \(s+1\). For each \(i\in\{1,\ldots,s\}\), write \(u_i=x_1\cdots x_i, \qquad v_i=x_{i+1}\cdots x_{s+1}.\) Then \(x=u_iv_i,\) and the corresponding ligation \(r_i:u_i+v_i\longrightarrow x\) is a forward reaction in \(R_n^+\). Moreover,

	$$
|u_i|\le s \qquad\text{and}\qquad |v_i|\le s,
$$

	so both reactants belong to \(X(s)\). Thus there are \(s\) distinct forward reactions \(r_1,\ldots,r_s\)
	that can produce \(x\) from molecules already generated at earlier stages. Consequently, \(x\) fails to be generated by a globally catalysed forward reaction only if all of the reactions \(r_1,\ldots,r_s\)
	are globally uncatalysed. For distinct reactions, the events of being globally uncatalysed are independent. Indeed, for each \(r\), this event depends only on the family \(\{I_{y,r}:y\in X_n\},\) and these families are disjoint for distinct reactions. Therefore

	$$
\mathbb P( r_1,\ldots,r_s \text{ are all globally uncatalysed} ) = \prod_{i=1}^{s}q_n(r_i).
$$

	Since \(q_n(r_i)\le q\) for every \(i\),

	$$
\mathbb P( x\text{ fails to be generated at level }s+1 ) \le q^s.
$$

	There are exactly \(\kappa^{s+1}\)
	molecules of length \(s+1\). Hence, by the union bound,

	$$
\mathbb P( \text{some molecule of length }s+1 \text{ fails to be generated} ) \le \kappa^{s+1}q^s.
$$

	Equivalently,

	$$
\mathbb P( \text{failure at level }s+1 ) \le \kappa(\kappa q)^s.
$$

	Applying the union bound over \(s=t,\ldots,n-1,\) we obtain

	$$
\mathbb P( \text{the recursive construction fails at some level} ) \le \kappa \sum_{s=t}^{n-1}(\kappa q)^s.
$$

	Since \(\kappa q<1\),

	$$
\kappa \sum_{s=t}^{n-1}(\kappa q)^s \le \kappa \sum_{s=t}^{\infty}(\kappa q)^s = \frac{ \kappa(\kappa q)^t }{ 1-\kappa q }.
$$

	It remains to verify that successful completion of this construction implies the existence of a RAF. Suppose that the construction succeeds at every level. For each molecule \(x\in X_n\setminus F,\) choose one globally catalysed forward reaction \(r_x:u+v\longrightarrow x\) used to generate \(x\), and let \(R' = \{r_x:x\in X_n\setminus F\}.\) By construction, the reactants of each selected reaction producing a molecule of length \(s+1\) have length at most \(s\). It follows inductively that every reactant of every reaction in \(R'\) can be generated from \(F\) using reactions in \(R'\). Thus \(R'\) is \(F\)-generated and \(\operatorname{cl}_{R'}(F)=X_n.\) Furthermore, every reaction in \(R'\) is globally catalysed. Hence, for each \(r\in R'\), there exists some molecule \(x\in X_n=\operatorname{cl}_{R'}(F)\) that catalyses \(r\). Therefore \(R'\) is reflexively autocatalytic. Since \(R'\) is both reflexively autocatalytic and \(F\)-generated, it is a RAF. Consequently,
\[
\mathbb P_n(\mathsf{RAF})\ge 1- \frac{ \kappa(\kappa q)^t }{ 1-\kappa q }.
\]

	Substituting \(q=e^{-s_-\lambda}\) gives

	$$
\mathbb P_n(\mathsf{RAF}) \ge 1- \frac{ \kappa\left(\kappa e^{-s_-\lambda}\right)^t }{ 1-\kappa e^{-s_-\lambda} }.
$$

	Finally, since \(t\), \(\kappa\), and \(s_->0\) are fixed, \(\kappa e^{-s_-\lambda}\longrightarrow0\) as \(\lambda\to\infty\). Therefore

	$$
\frac{ \kappa\left(\kappa e^{-s_-\lambda}\right)^t }{ 1-\kappa e^{-s_-\lambda} } \longrightarrow0,
$$

	and the stated lower bound converges to \(1\).
\end{proof}

\subsection{Proof of Proposition~\ref{prop:failure-MS-criteria}}

\begin{proof}
We construct the two sequences of kernels explicitly.
Throughout, we take $\lambda_n=1$.

We first prove (i). Fix two food molecules $a,b\in F$ such that their
concatenation $y=ab$ does not belong to $F$.
For example, we may choose $|a|=t$ and $|b|=1$,
so that $|y|=t+1$. Let
\[
r^\ast:a+b\longrightarrow y
\]
be the corresponding forward ligation reaction.

For every $n\ge t+1$, define
\[
K_n(x,r)
=
\begin{cases}
n, & x=y,\quad r=r^\ast,\\
0, & \text{otherwise}.
\end{cases}
\]
Consequently, the only catalysis event with positive
probability is that $y$ catalyses $r^\ast$, and
\[
\mathbb P(I_{y,r^\ast}=1)=1-e^{-n}.
\]

For each molecule $x\in X_n$, the expected number of
reactions catalysed by $x$ is therefore
\[
\mu_n(x)
=
\begin{cases}
1-e^{-n}, & x=y,\\
0, & x\ne y.
\end{cases}
\]
It follows that
\[
0\le
\frac{1}{n}\max_{x\in X_n}\mu_n(x)
=
\frac{1-e^{-n}}{n}
\le \frac1n
\longrightarrow0.
\]

On the other hand, consider the singleton reaction set
\[
R'=\{r^\ast\}.
\]
Its reactants $a$ and $b$ belong to $F$, and hence
$R'$ is $F$-generated. Moreover, its product $y$
belongs to the closure of $F$ under $R'$.
Whenever $I_{y,r^\ast}=1$, the reaction $r^\ast$
is catalysed by $y$, which belongs to this closure.
Thus $R'$ is a RAF on the event
$\{I_{y,r^\ast}=1\}$.

Conversely, if $I_{y,r^\ast}=0$, no reaction in
$R_n^+$ is catalysed, so no nonempty RAF can exist.
Therefore
\[
\mathbb P_n(\mathsf{RAF})
=
\mathbb P(I_{y,r^\ast}=1)
=
1-e^{-n}
\longrightarrow1.
\]
This proves (i).

It remains to prove (ii).
Let
\[
\mathcal B_t
=
\{r\in R_n^+:\rho(r)\subseteq F\}
\]
denote the set of forward reactions whose reactants
both belong to the food set. Since $t$ is fixed,
the cardinality
 $
B_t:=|\mathcal B_t|
$
 is finite and independent of $n$ for all sufficiently
large $n$.

Fix $p\in(0,1)$ and put
\[
c_p=-\log(1-p)>0.
\]
Define
\[
K_n(x,r)
=
\begin{cases}
0, & r\in\mathcal B_t,\\
c_p, & r\in R_n^+\setminus\mathcal B_t,
\end{cases}
\qquad x\in X_n.
\]
It follows that
\[
\mathbb P(I_{x,r}=1)
=
\begin{cases}
0, & r\in\mathcal B_t,\\
p, & r\in R_n^+\setminus\mathcal B_t.
\end{cases}
\]
In particular, for every molecule $x\in X_n$,
\[
\mu_n(x)
=
p\bigl(|R_n^+|-B_t\bigr).
\]
Since $|x|\le n$, we obtain
\[
\min_{x\in X_n}
\frac{\mu_n(x)}{|x|}
=
\frac{p(|R_n^+|-B_t)}{n}.
\]
We now recall that the number of forward ligation
reactions is
$
|R_n^+|
=
\displaystyle \sum_{\ell=2}^n(\ell-1)\kappa^\ell$.
In particular,
\[
|R_n^+|
\ge (n-1)\kappa^n,
\]
and therefore
\[ 
\dln
\min_{x\in X_n}
\frac{\mu_n(x)}{|x|}
\ge
\dln p\left(
\frac{n-1}{n}\kappa^n
-\frac{B_t}{n}
\right)
=\infty.
\]
Nevertheless, by construction, no reaction
$r\in\mathcal B_t$ can be catalysed.

We claim that every nonempty $F$-generated reaction
set contains at least one reaction from $\mathcal B_t$.
Indeed, suppose that $R'$ is a nonempty
$F$-generated reaction set. By the definition of
$F$-generation, its reactions admit an ordering
$r_1,\ldots,r_m$ such that the reactants of $r_1$
belong to $F$, and the reactants of each subsequent
reaction belong to the food set or are products of
preceding reactions. Hence $r_1\in\mathcal B_t$.

In particular, every RAF contains a reaction from
$\mathcal B_t$. Since no such reaction is catalysed,
no RAF can exist. Consequently,
\[
\mathbb P_n(\mathsf{RAF})=0
\]
for all sufficiently large $n$. This proves (ii) and completes the proof.
\end{proof}

\subsection{Proof of Theorem~\ref{thm:weighted-mass-criterion}}

\begin{proof}
	For the first statement, every nontrivial RAF must contain at least one
	reaction whose reactants are contained in the food set. Consequently,
	conditional on the weights,
	\[
\mathbb P_n(\mathrm{RAF}\mid (W_x)) \leq \sum_{r\in\mathcal B_t} \left(1-e^{-\lambda_nM_n(r)}\right).
\]
	Writing \(Z_n = \lambda_n \max_{r\in\mathcal B_t}M_n(r),\)
	we obtain 
\[
\mathbb P_n(\mathrm{RAF}\mid (W_x)) \leq |\mathcal B_t|(1-e^{-Z_n}).
\]

	By assumption \(Z_n\to0\) in probability. Since \(0\leq1-e^{-Z_n}\leq1\), it follows that
\(\mathbb E[1-e^{-Z_n}]\longrightarrow0.\)
	As \(|\mathcal B_t|\) is bounded independently of \(n\), averaging over
	the weights gives \(\mathbb P_n(\mathrm{RAF})\longrightarrow0.\) For the second statement, put \(Y_n = \lambda_n \min_{r\in R_n^+}M_n(r).\)
	Conditional on the weights, every reaction is uncatalysed with
	probability at most \(e^{-Y_n}.\)
	Furthermore, the uncatalysed events of distinct reactions are
	conditionally independent. Fix \(b>\log\kappa\). On the event \(Y_n\geq b\), every reaction is
	uncatalysed with conditional probability at most \(e^{-b}\). The usual
	recursive polymer construction therefore gives
	\[
\mathbb P_n(\mathrm{RAF}^c\mid (W_x)) \leq \frac{ \kappa(\kappa e^{-b})^t }{ 1-\kappa e^{-b} }
\]
	on this event. Hence
	\[
\mathbb P_n(\mathrm{RAF}^c) \leq \mathbb P(Y_n<b) + \frac{ \kappa(\kappa e^{-b})^t }{ 1-\kappa e^{-b} }.
\]
	Since \(Y_n\to\infty\) in probability,
	\[
\limsup_{n\to\infty}\mathbb P_n(\mathrm{RAF}^c) \leq \frac{ \kappa(\kappa e^{-b})^t }{ 1-\kappa e^{-b} }.
\]
	Letting \(b\to\infty\) shows that the right-hand side tends to zero.
	Thus \(\mathbb P_n(\mathrm{RAF})\longrightarrow1.\)
\end{proof}

\subsection{Proof of Proposition~\ref{prop:finitemean}}\label{sec:finitemean proof}

\begin{proof}
	For \(r\in\mathcal B_t\), \(\mathbb E[M_n(r)] = \frac{\alpha}{\alpha-1}S_n(r).\)
	Thus, for every \(\varepsilon>0\),
	\[
\mathbb P\bigl(\lambda_nM_n(r)>\varepsilon\bigr) \leq \frac{\alpha}{\alpha-1} \frac{\lambda_nS_n(r)}{\varepsilon}.
\]
	Since \(|\mathcal B_t|\) is bounded, 
\[
\lambda_n \max_{r\in\mathcal B_t}M_n(r) \xrightarrow{\mathbb P}0.
\]

	The first assertion follows from Theorem~\ref{thm:weighted-mass-criterion}. For the second assertion, Pareto weights satisfy \(W_x\geq1\) almost
	surely. Hence \(M_n(r) = \sum_xW_xK_n(x,r) \geq S_n(r) \geq s_-.\)
	Therefore \(\lambda_n \min_{r\in R_n^+}M_n(r) \geq s_-\lambda_n \longrightarrow\infty,\)
	and the second part of Theorem~\ref{thm:weighted-mass-criterion} applies.
\end{proof}

The same argument also recovers a useful fixed-\(\lambda\) lower bound.
If \(S_n(r)\geq s_->0\) for every reaction and \(\lambda>\frac{\log\kappa}{s_-},\)
then
\[
\mathbb P_n(\mathsf{RAF}) \geq 1- \frac{ \kappa(\kappa e^{-s_-\lambda})^t }{ 1-\kappa e^{-s_-\lambda} }.
\]
Thus the finite-mean regime retains the ordinary catalytic-intensity
scale: decreasing \(\lambda_n\) to zero suppresses RAFs, while increasing
the global catalytic intensity produces RAFs with high probability.

\subsection{Proof of Theorem~\ref{thn:infcrit}}

\begin{proof}
	Since \(\sup_nS_n<\infty\) and \(\lambda_n\to0\), \(\lambda_n\max_x k_n(x) \leq \lambda_n S_n \longrightarrow0.\)
	Thus, for all sufficiently large \(n\), \(\lambda_nk_n(x)\leq1\)
	for every \(x\). We first prove the subcritical statement. Fix \(0<\varepsilon<1\) and
	put \(z_n=\frac{\varepsilon}{\lambda_n}.\)
	Write
	\[
M_n = \sum_x k_n(x)W_x \mathbf 1_{\{k_n(x)W_x\leq z_n\}} + \sum_x k_n(x)W_x \mathbf 1_{\{k_n(x)W_x>z_n\}}.
\]
	Therefore \(\mathbb P(M_n>2z_n)\)
	is bounded by
	\[
\mathbb P\left( \exists x:\ k_n(x)W_x>z_n \right) + \mathbb P\left( \sum_x k_n(x)W_x \mathbf 1_{\{k_n(x)W_x\leq z_n\}} >z_n \right).
\]

	Since \(W\) has Pareto tail \(w^{-1}\), 
\[
\mathbb P\bigl(k_n(x)W_x>z_n\bigr) = \frac{k_n(x)}{z_n}
\]

	for all sufficiently large \(n\). Hence
	\[
\mathbb P\left( \exists x:\ k_n(x)W_x>z_n \right) \leq \frac{\lambda_nS_n}{\varepsilon}.
\]

	For the truncated contribution,
	\[
\mathbb E\left[ k_n(x)W_x \mathbf 1_{\{k_n(x)W_x\leq z_n\}} \right] = k_n(x) \log\frac{z_n}{k_n(x)}.
\]
	Consequently, by Markov's inequality,
	\[
\mathbb P\left( \sum_x k_n(x)W_x \mathbf 1_{\{k_n(x)W_x\leq z_n\}} >z_n \right) \leq \frac{\lambda_n}{\varepsilon} \sum_x k_n(x) \log\frac{\varepsilon}{\lambda_nk_n(x)}.
\]
	Since \(0<\varepsilon<1\), 
\[
\log\frac{\varepsilon}{\lambda_nk_n(x)} \leq \log\frac{1}{\lambda_nk_n(x)},
\]

	and therefore
	\[
\mathbb P(\lambda_nM_n>2\varepsilon) \leq \frac{\lambda_nS_n}{\varepsilon} + \frac{A_n^{(1)}}{\varepsilon}.
\]
	If \(A_n^{(1)}\to0\), then the right-hand side tends to zero because \(\lambda_nS_n\to0\). Thus
\(\lambda_nM_n\xrightarrow{\mathbb P}0.\)
	As all reactions have the same mass \(M_n\), the first part of Theorem~\ref{thm:weighted-mass-criterion} gives \(\mathbb P_n(\mathsf{RAF})\longrightarrow0.\) For the supercritical statement, recall that for Pareto weights with \(\alpha=1\),
\(\mathbb P(W>w)=w^{-1}, \qquad w\geq1.\)
	Let
	\[
\phi_1(u) := \mathbb E[e^{-uW}] = \int_1^\infty e^{-uw}w^{-2}\,dw, \qquad u>0,
\]
	denote the Laplace transform of \(W\). Then, for \(0<u\leq1\),
	\[
1-\phi_1(u) = \mathbb E[1-e^{-uW}] \geq \mathbb E\left[ (1-e^{-uW}) \mathbf 1_{\{1\leq W\leq 1/u\}} \right].
\]
	For \(1\leq W\leq1/u\), we have \(0\leq uW\leq1\), and hence \(1-e^{-uW}\geq e^{-1}uW.\)
	Therefore
	\[
\begin{aligned} 1-\phi_1(u) \geq e^{-1}u\, \mathbb E\left[ W\mathbf 1_{\{1\leq W\leq1/u\}} \right] = e^{-1}u \int_1^{1/u} w\,w^{-2}\,dw= \frac{u}{e}\log\frac1u.
\end{aligned}
	\]
	Thus 
\[
1-\phi_1(u) \geq \frac{u}{e}\log\frac1u, \qquad 0<u\leq1.
\]

	Since \(\log z\leq z-1\) for $z>04$,
	we have \(\log\phi_1(u) \leq -\bigl(1-\phi_1(u)\bigr),\)
	and consequently
	\[
\phi_1(u) \leq \exp\left\{ -\frac{u}{e}\log\frac1u \right\}, \qquad 0<u\leq1.
\]

	For any \(m>0\), exponential Markov inequality gives 
\[
\mathbb P(M_n\leq m) \leq e^{\lambda_nm} \mathbb E[e^{-\lambda_nM_n}].
\]

	Independence of the weights yields 
\[
\mathbb E[e^{-\lambda_nM_n}] = \prod_x \phi_1(\lambda_nk_n(x)).
\]

	Hence
	\[
\mathbb P(M_n\leq m) \leq \exp\left\{ \lambda_nm - \frac{\lambda_n}{e} J_n(\lambda_n) \right\}.
\]
	Choose \(m= \frac{1}{2e}J_n(\lambda_n).\)
	Then
	\[
\mathbb P\left( M_n\leq \frac{1}{2e}J_n(\lambda_n) \right) \leq \exp\left\{ -\frac{A_n^{(1)}}{2e} \right\}.
\]
	If \(A_n^{(1)}\to\infty\), the right-hand side tends to zero. Thus, with
	probability tending to one, \(\lambda_nM_n \geq \frac{A_n^{(1)}}{2e},\)
	and consequently \(\lambda_nM_n\xrightarrow{\mathbb P}\infty.\)
	The second part of Theorem~\ref{thm:weighted-mass-criterion} now gives \(\mathbb P_n(\mathsf{RAF})\longrightarrow1.\)
\end{proof}

Thus, outside the critical window \(A_n^{(1)}=O(1)\), the critical
infinite-mean model exhibits a clean phase separation governed by the
single quantity \(A_n^{(1)} = \lambda_n \sum_x k_n(x) \log\frac{1}{\lambda_nk_n(x)}.\)

\subsection{Proof of Theorem~\ref{thm:inf}}

\begin{proof}
	We first establish the subcritical implication. Put \(M_n=\sum_xk_n(x)W_x.\)
	Fix \(\varepsilon>0\) and let \(z_n=\frac{\varepsilon}{\lambda_n}.\)
	If \(M_n>2z_n\), then either some individual contribution exceeds \(z_n\), or the sum of the contributions truncated at \(z_n\) exceeds
	\(z_n\). Hence
	\[
\mathbb P(M_n>2z_n) \leq \mathbb P\left( \exists x:\ k_n(x)W_x>z_n \right) + \mathbb P\left( \sum_x k_n(x)W_x \mathbf 1_{\{k_n(x)W_x\leq z_n\}} >z_n \right).
\]

	For the first term, 
\[
\mathbb P(k_n(x)W_x>z_n) \leq \left(\frac{k_n(x)}{z_n}\right)^\alpha.
\]

	This inequality also holds when \(k_n(x)>z_n\), since then the
	right-hand side is at least one. Therefore
	\[
\mathbb P\left( \exists x:\ k_n(x)W_x>z_n \right) \leq z_n^{-\alpha} \sum_xk_n(x)^\alpha.
\]

	For the truncated contribution, if \(k_n(x)\leq z_n\), then
	\[
\begin{aligned} \mathbb E\left[ k_n(x)W_x \mathbf 1_{\{k_n(x)W_x\leq z_n\}} \right] = \alpha k_n(x) \int_1^{z_n/k_n(x)} w^{-\alpha}\,dw \leq \frac{\alpha}{1-\alpha} k_n(x)^\alpha z_n^{1-\alpha}.
\end{aligned}
	\]
	If \(k_n(x)>z_n\), the truncated random variable is identically zero,
	because \(W_x\geq1\). Markov's inequality therefore gives
	\[
\mathbb P\left( \sum_x k_n(x)W_x \mathbf 1_{\{k_n(x)W_x\leq z_n\}} >z_n \right) \leq \frac{\alpha}{1-\alpha} z_n^{-\alpha} \sum_xk_n(x)^\alpha.
\]
	Combining the two estimates,
	\[
\mathbb P(M_n>2z_n) \leq \frac{1}{1-\alpha} z_n^{-\alpha} \sum_xk_n(x)^\alpha.
\]
	Since \(z_n=\varepsilon/\lambda_n\),
	\[
\mathbb P(\lambda_nM_n>2\varepsilon) \leq \frac{1}{(1-\alpha)\varepsilon^\alpha} A_n^{(\alpha)}.
\]
	Thus \(A_n^{(\alpha)}\to0\)
	implies \(\lambda_nM_n\xrightarrow{\mathbb P}0.\)
	Theorem~\ref{thm:weighted-mass-criterion} gives \(\mathbb P_n(\mathrm{RAF})\longrightarrow0.\) For the supercritical implication, use \(M_n \geq \max_x\{k_n(x)W_x\}.\)
	Fix \(B>0\). Then
	\[
\mathbb P(\lambda_nM_n\leq B) \leq \mathbb P\left( \lambda_nk_n(x)W_x\leq B \text{ for every }x \right).
\]

	If \(\lambda_nk_n(x)>B\) for some \(x\), then, since \(W_x\geq1\), \(\lambda_nk_n(x)W_x>B\)
	almost surely, and hence \(\mathbb P(\lambda_nM_n\leq B)=0.\) Otherwise, \(\lambda_nk_n(x)\leq B\)
	for every \(x\). By the Pareto tail formula,
	\[
\mathbb P\left( W_x\leq \frac{B}{\lambda_nk_n(x)} \right) = 1- \left( \frac{\lambda_nk_n(x)}{B} \right)^\alpha.
\]
	Independence gives
	\[
\mathbb P(\lambda_nM_n\leq B) \leq \prod_x \left[ 1- \left( \frac{\lambda_nk_n(x)}{B} \right)^\alpha \right].
\]
	Using \(1-u\leq e^{-u}\),
	\[
\mathbb P(\lambda_nM_n\leq B) \leq \exp\left\{ -\frac{\lambda_n^\alpha}{B^\alpha} \sum_xk_n(x)^\alpha \right\}.
\]
	Thus
	\[
\mathbb P(\lambda_nM_n\leq B) \leq \exp\left\{ -\frac{A_n^{(\alpha)}}{B^\alpha} \right\}.
\]
	If \(A_n^{(\alpha)}\to\infty\), this tends to zero for every fixed \(B>0\). Therefore
\(\lambda_nM_n\xrightarrow{\mathbb P}\infty.\)
	Theorem~\ref{thm:weighted-mass-criterion} now implies \(\mathbb P_n(\mathrm{RAF})\longrightarrow1.\)
\end{proof}

Thus, for \(0<\alpha<1\), the phase transition is governed by \(A_n^{(\alpha)} = \lambda_n^\alpha \sum_xk_n(x)^\alpha.\)
Outside the critical window \(A_n^{(\alpha)}=O(1)\), \(A_n^{(\alpha)}\to0 \quad\Longrightarrow\quad\mathbb P_n(\mathsf{RAF})\to0,\)
whereas
\[
A_n^{(\alpha)}\to\infty \quad\Longrightarrow\quad\mathbb P_n(\mathsf{RAF})\to1.
\]

\subsection{Proof of Proposition~\ref{prop:single-catalyst-raf}}
\begin{proof}
Let
\[
W_n^{\max}=\max_{x\in X_n}W_x,
\]
and choose a molecule $x_n^\ast\in X_n$ attaining
this maximum.

Since the weights are i.i.d.\ Pareto-distributed,
for every $u\ge1$,
\[
\mathbb P(W_n^{\max}\le u)
=
(1-u^{-\alpha})^{N_n}
\le \exp(-N_nu^{-\alpha}).
\]

Set
\[
Z_n=\frac{\lambda_n W_n^{\max}}{N_n}.
\]
We first show that
\[
Z_n-\log N_n\xrightarrow{\mathbb P}\infty.
\]
Indeed, for every fixed $A>0$, and all sufficiently
large $n$,
\begin{align*}
\mathbb P(Z_n\le \log N_n+A)
=
\mathbb P\left(
W_n^{\max}\le
\frac{N_n(\log N_n+A)}{\lambda_n}
\right)\le
\exp\left(
-\frac{\lambda_n^\alpha N_n^{1-\alpha}}
{(\log N_n+A)^\alpha}
\right)
\to, \text{as $n\to \infty$},
\end{align*}
where the convergence follows from the assumption.

Next, choose a deterministic set of forward ligation
reactions $\mathcal R_n\subseteq R_n^+$ that generates
every molecule in $X_n$ from $F$.
Such a set can be constructed by selecting, for each
molecule $y\in X_n\setminus F$, one forward ligation
reaction producing $y$ from shorter molecules.
Consequently,
\[
|\mathcal R_n|=N_n-|F|\le N_n,
\]
and
\[
\operatorname{cl}_{\mathcal R_n}(F)=X_n.
\]

Conditional on the weights, the probability that
$x_n^\ast$ does not catalyse a given reaction
$r\in\mathcal R_n$ is
\[
\mathbb P(I_{x_n^\ast,r}=0\mid W)
=
e^{-Z_n}.
\]
Therefore, by the union bound,
\[
\mathbb P\left(
\exists r\in\mathcal R_n:
I_{x_n^\ast,r}=0
\,\middle|\,W
\right)
\le N_ne^{-Z_n}.
\]

Since
\[
Z_n-\log N_n\xrightarrow{\mathbb P}\infty,
\]
we have
\[
\min\{1,N_ne^{-Z_n}\}
\xrightarrow{\mathbb P}0.
\]
As this random variable is bounded by one,
its expectation also tends to zero. Hence
\[
\mathbb P\left(
I_{x_n^\ast,r}=1
\text{ for every }r\in\mathcal R_n
\right)
\longrightarrow1.
\]

On this event, every reaction in $\mathcal R_n$
is catalysed by the same molecule $x_n^\ast$.
Moreover, $\mathcal R_n$ is $F$-generated and
\[
x_n^\ast\in
\operatorname{cl}_{\mathcal R_n}(F)=X_n.
\]
Thus $\mathcal R_n$ is a nonempty RAF whose
every reaction is catalysed by $x_n^\ast$.

It follows that
\[
\mathbb P_n(\mathsf{RAF})\longrightarrow1.
\]
This completes the proof.
\end{proof}

\section{Discussion}

In this paper, we investigated the probability of RAF formation in random catalytic reaction systems with random molecular weights and reaction-dependent catalysis. We established general criteria for RAF formation in terms of weighted catalytic masses $M_n$ and the intensity $\lambda_n$. We further showed that classical criteria based on the expected number of reactions catalyzed by each molecule need not hold when catalysis is reaction-dependent. For heavy-tailed molecular weights, we identified distinct phase-transition regimes depending on the tail index. In particular, we proved that RAFs may emerge with probability tending to one even when the global catalytic intensity vanishes (e.g. Theorem \ref{thm:weighted-mass-criterion} and  Proposition \ref{prop:failure-MS-criteria}) . Moreover, under suitable conditions, a single exceptionally active molecule can catalyze all reactions of a RAF with high probability (Proposition \ref{prop:single-catalyst-raf}). These results suggest that the emergence of self-sustaining catalytic systems does not necessarily require catalytic activity to be broadly distributed among molecules. Instead, a small number of highly active molecules may play a decisive role in RAF formation.

An interesting direction for future research is to connect the probabilistic theory of RAF formation with the mathematical theory of chemical reaction networks (CRNs). A recent study by Golnik et al. \cite{golnik2026bridging} established an important connection between catalytic reaction systems, which is considered in the present paper, and CRNs by relating reflexive autocatalysis to stoichiometric autocatalysis. Their subsequent work \cite{golnik2026minimality} further investigated the relationship between irreducible RAFs and autocatalytic cores, while related studies developed methods for identifying autocatalytic subsystems in large CRNs \cite{golnik2026enumeration,golnik2026using}. These developments suggest a natural extension of our work: investigating the probability of RAF formation in random CRNs. Unlike the present model, where the set of possible reactions is fixed and only catalytic relationships are random, such a framework would allow the reaction network itself to be random. It would be interesting to characterize the probability of RAF formation and identify possible phase transitions as the reaction network structure and catalytic parameters vary. In particular, one could investigate how phase transitions for RAF formation are related to those for stoichiometric autocatalysis.

Experimental studies have also demonstrated the emergence and organization of autocatalytic chemical systems, including cooperative RNA replication networks \cite{vaidya2012spontaneous}, small-molecule autocatalysis coupled to compartment growth and competition \cite{lu2024small}, and programmable autocatalytic reaction networks \cite{kriukov2024exploring}. These studies highlight the potential role of autocatalytic interactions in the emergence of organized chemical behavior. Our results provide a complementary theoretical perspective by showing that strong heterogeneity in molecular catalytic activity can substantially change the conditions under which autocatalytic networks emerge. In particular, the possibility that a small number of highly active molecules can support a RAF suggests an interesting question for future experimental investigations: whether similarly concentrated catalytic activity can facilitate the emergence of self-sustaining chemical networks. Connecting such experimental observations with probabilistic models of RAF formation may help clarify how heterogeneous catalytic interactions contribute to the emergence of chemical organization relevant to the origin of life.



\bigskip
\noindent\textbf{Declaration of AI use.}
ChatGPT (OpenAI) was used to assist with mathematical discussions,
drafting, searching references, writing a code for Figure \ref{fig:raf samples}, and language editing. The authors take full responsibility
for the content of the manuscript.
\\

\noindent\textbf{Acknowledgments.}
JK was supported by the National Research Foundation of Korea (NRF) grant funded by the Korea government(MSIT) (RS-2023-00219980), and Korea Health Technology R\&D Project through the Korea Health Industry Development Institute (KHIDI), funded by the Korea government (MSIT) (No. RS-2024-00406488). JK and ES were supported by National Research Foundation of Korea (NRF) grant funded by the Korea government (MSIT) (No. RS-2026-25486346 and No.2021R1A6A1A10042944)

\end{document}